\documentclass[a4paper,11pt]{article}
\pdfoutput=1
\usepackage{jcappub}

\usepackage{MnSymbol}
\usepackage[applemac]{inputenc}
\usepackage{graphicx}
\usepackage{amsmath}
\usepackage{amsfonts}
\usepackage{amssymb}
\usepackage{makeidx}
\usepackage{wrapfig}
\usepackage{empheq}
\usepackage{geometry}
\usepackage{comment}
\usepackage{soul}
\usepackage{ulem}

\newcommand{\bk}{{\mathbf k}}
\newcommand{\bq}{{\mathbf q}}

\newcommand{\bx}{{\mathbf x}}

\newcommand{\bn}{{\mathbf n}}
\newcommand{\bv}{{\mathbf v}}
\newcommand{\bnabla}{{\mathbf \nabla}}

\newcommand{\De}{\Delta}

\newcommand{\ga}{\gamma}

\newcommand{\ka}{\kappa}

\newcommand{\La}{\Lambda}

\newcommand{\Om}{\Omega}

\newcommand{\si}{\sigma}

\newcommand{\be}{\begin{equation}}
\newcommand{\ee}{\end{equation}}

\newcommand{\bea}{\begin{eqnarray}}
\newcommand{\eea}{\end{eqnarray}}
\newcommand{\bean}{\begin{eqnarray*}}
\newcommand{\eean}{\end{eqnarray*}}

\newcommand{\HH}{{\cal H}}
\newcommand{\CHI}{S_K}

\newcommand{\class}{{\sc class}}
\newcommand{\classgal}{{\sc class}gal}

\usepackage{color}
\usepackage[usenames,dvipsnames,svgnames,table]{xcolor}

\definecolor{darkred}{RGB}{175,0,0}
\definecolor{darkblue}{RGB}{14,0,185}
\definecolor{grey}{RGB}{0,139,139}
\definecolor{dgreen}{rgb}{0,0.6,0} 

\title{Curvature constraints from Large Scale Structure}

\author[a,b]{Enea~Di~Dio}
\author[c,d]{Francesco~Montanari}
\author[e]{Alvise~Raccanelli}
\author[d]{Ruth~Durrer}
\author[e]{Marc~Kamionkowski}
\author[f]{Julien~Lesgourgues}

\affiliation[a]{INAF - Osservatorio Astronomico di Trieste, Via G. B. Tiepolo 11, I-34143 Trieste, Italy}
\affiliation[b]{INFN, Sezione di Trieste, Via Valerio 2, I-34127 Trieste, Italy}
\affiliation[c]{Physics Department, University of Helsinki and Helsinki Institute of Physics, P.O. Box 64, FIN-00014, University of Helsinki, Finland
}
\affiliation[d]{D\'epartement de Physique Th\'eorique and Center for Astroparticle Physics, Universit\'e de Gen\`eve, 24 quai Ernest Ansermet, CH--1211 Gen\`eve 4, Switzerland}
\affiliation[e]{Department of Physics \& Astronomy, Johns Hopkins University, Baltimore, MD 21218, USA}
\affiliation[f]{Institut f\"ur Theoretische Teilchenphysik und Kosmologie, RWTH Aachen University, D-52056 Aachen, Germany}

\emailAdd{enea.didio@oats.inaf.it}
\emailAdd{francesco.montanari@helsinki.fi}
\emailAdd{alvise@jhu.edu}
\emailAdd{ruth.durrer@unige.ch}
\emailAdd{kamion@jhu.edu}
\emailAdd{lesgourg@physik.rwth-aachen.de}


\abstract{
We modified the \class{} code in order to include relativistic galaxy number counts in spatially curved geometries; we present the formalism and study the effect of relativistic corrections on spatial curvature. The new version of the code is now publicly available.
Using a Fisher matrix analysis, we investigate how measurements of the spatial curvature parameter $\Omega_K$ with future galaxy surveys are affected by relativistic effects, which influence observations of the large scale galaxy distribution. These effects  include contributions from cosmic magnification, Doppler terms and terms involving the gravitational potential. As an application, we consider angle and redshift dependent power spectra, which are especially well suited for model independent cosmological constraints.
We compute our results for a representative deep, wide and spectroscopic survey, and our results show the impact of relativistic corrections on spatial curvature parameter estimation.
We show that constraints on the curvature parameter may be strongly biased if, in particular, cosmic magnification is not included in the analysis.
Other relativistic effects turn out to be subdominant in the studied configuration.
We analyze how the shift in the estimated best-fit value for the curvature and other cosmological parameters depends on the magnification bias parameter, and find that significant biases are to be expected if this term is not properly considered in the analysis.
}

\begin{document}
\maketitle
\flushbottom

\section{Introduction}
\label{sec:intro}
Spatial curvature is one of the most interesting parameters to measure in order to constrain properties of the early universe.
Important constraints on the following parameters have been set to describe different inflationary models in the current picture of the primordial universe: the deviation from Gaussianity in the primordial perturbations, quantified, e.g., by the primordial bispectrum amplitude parameter $f_{\rm NL}$ and its running $n_{\rm NG}$, the tensor-to-scalar ratio $r$, the spectral index of the primordial power spectrum $n_s$ and its running $\alpha_s$, and the curvature parameter $\Omega_{\rm K}$. The ability to constrain the first two of these parameters has been investigated in several works by looking at both the CMB~\citep[see e.g.][]{Bartolo:2004, Komatsu:2010, planck2015nG} and the large-scale structure (LSS) of the universe~\citep[see e.g.][]{Matarrese:2000, Dalal:2008, Matarrese:2008, Raccanelli:2014fNL, dePutter:2014, Alvarez:2014, Raccanelli:2015GR}, or a combination of the two~\citep[see e.g.][]{Xia:2010, Mao:2013, Raccanelli:2014isw}. The parameter $r$ is usually constrained using CMB experiments~\citep[see e.g.][]{bicep2, planckbicep2, planck2015nG, planck2015inf}.
The primordial power spectrum parameters $\{n_s,\alpha_s\}$ are routinely used to test cosmological models~\citep[see e.g.][]{wmap, Planck:2015xua, Sanchez:2012}.

The importance of measuring curvature as a means to constrain inflationary models has been discussed, e.g., in~\cite{Linde:2003, Linde:2014, Guth:2012, Guth:2013}. Detecting $\Omega_K\neq 0$ would significantly limit the parameter space for inflationary models. One of the primary motivations for inflationary cosmology was to explain the observed (spatial) flatness of our observed Universe.
Inflationary models that allow a large number of e-foldings predict that our Universe should be very accurately spatially flat, and the effective curvature within our Hubble radius should be of the order of the amplitude of the curvature fluctuations generated during inflation, i.e.~$\Omega_K \sim \mathcal{O}(10^{-5}$).

Inflationary models with interesting open~\citep[see e.g.][]{Gott:1982, Kamionkowski:1994qf, Kamionkowski:1994xy, Linde:1996, Bucher:1994, Linde:1998} or closed geometries~\cite{Linde:2003} have been considered. Even more speculatively, models with open geometries from tunneling events between metastable vacua within a ``string landscape'' have been proposed~\cite{Freivogel:2006}. Observational limits on spatial curvature therefore offer important additional constraints on inflationary models and fundamental physics.
It has been shown that the most robust implication of false vacuum decay is negative curvature~\cite{Freivogel:2006}, which can naturally arise in the multiverse, while it would be difficult to understand it within standard cosmological models without significant fine tuning. Interestingly, a discovery of positive curvature would falsify the multiverse as we understand it. Furthermore, the exploration of curvature may contain information about possible measures in the multiverse~\cite{Guth:2012}.

Let us also note that departure from flatness could be due either to a local inhomogeneity, or a truly superhorizon departure from flatness. In the first case, spectral distortions can be induced at a level detectable by next-generation experiments \cite{Bull:2013fga}.

Current measurements of the curvature parameter come from the combination of CMB~\cite{Kamionkowski:1994nr, Jungman:1996rm} and LSS~\cite{Takada:2015db}.
The SPT lensing measurements combined with 7y WMAP temperature spectrum found $\Omega_K  = 0.0014 \pm 0.017$~\cite{Engelen:2012}.
The Planck satellite recently constrained the curvature using observations of the CMB alone, finding $\Omega_K = -0.040^{+0.038}_{-0.041}$; the addition of lensing reconstruction gives a substantial improvement, bringing the constraints to $\Omega_K = -0.005^{+0.016}_{-0.017}$.
These constraints are substantially improved by the addition of BAO data, and the combined constraints read $\Omega_K = -0.0001^{+0.0054}_{-0.0052}$~\cite{Planck:2015xua}.
The curvature parameter has also been measured by galaxy surveys such as BOSS, which arrived at $\Omega_K =-0.00264^{+0.00466}_{-0.00461}$, or $\Omega_K = -0.00411 \pm 0.01029$ with more conservative assumptions~\cite{Zhao:2012}, and with a more recent analysis $\Omega_K = 0.0010\pm0.0029$~\cite{Sanchez:2013}; WiggleZ found $\Omega_K = -0.0043 \pm 0.0047$~\cite{Kazin:2014}. Note that all the parameter estimations from LSS observations so far are based on a standard power spectrum, $P(k)$, analysis which does not take into account the relativistic effects studied in this work.
There are also prospects to learn about $\Omega_K$ from measurements of the jerk, the third derivative of the expansion rate. The jerk was first reported measured in~\cite{Riess:2004yg}, and the implications for $\Omega_K$ discussed in~\cite{Caldwell:2004lq}.

The current curvature parameter estimations show that we are not yet at the cosmic variance limit.
These measurements are more than one order of magnitude away from the limiting threshold to which we can test the models mentioned above.
  Furthermore, based on statistical arguments~\cite{Vardanyan:2009}, the information coming from joint CMB and planned future large scale structure surveys may not allow unambiguous conclusions about the geometry of our Universe if the value of the curvature parameter is below $|\Omega_K|\sim\mathcal{O}(10^{-4})$.

  In recent years there has been a significant increase in the amount and quality of the available cosmological data, and the next decade promises to bring further substantial improvements.
  It is important to understand how precisely future instruments will constrain the curvature parameter, trying to provide insights to some of the most profound questions regarding the primordial universe, and even providing observational constraints for string theory models.
  Future galaxy surveys will probe huge cosmological volumes by building wide~\citep[see e.g.][]{Maartens:2015mra, spherex} and deep~\citep[see e.g.][]{pfs, wfirst, Laureijs:2011gra} galaxy catalogs. It has been argued that a proper relativistic treatment will be needed in order to provide a sufficiently accurate theoretical modeling of galaxy clustering on those scales~\citep[see e.g.][]{Yoo:2009, Yoo:2010, Bonvin:2011bg,Challinor:2011bk, Bertacca:2012, Jeong:2012, Raccanelli:2013, Raccanelli:radial}. Recently the effects of relativistic corrections on 2-point~\cite{Lombriser:2013, Raccanelli:2012kq, Raccanelli:2013growth,Raccanelli:2015GR,Alonso:2015uua,Baker:2015bva,Montanari:2015rga,Fonseca:2015laa} and higher order statistics~\cite{DiDio:2014lka,DiDio:2015bua,Kehagias:2015tda} have been studied, and methods to isolate them have been proposed~\cite{McDonald:2009ud,Bonvin:2013ogt,Irsic:2015nla,Bonvin:2015kuc,Gaztanaga:2015jrs}.
It is now timely to investigate the potential of large-scale effects to
measurements of the curvature parameter. Given that a Newtonian description of galaxy clustering might lead to biased results, we pay special attention to the relativistic  effects.

In this paper we describe the modification of the CLASS code\footnote{\url{http://class-code.net/}}~\cite{Lesgourgues:2011re,Blas:2011rf,Lesgourgues:2013bra} needed to include spatially curved geometries not only in the computation of CMB observables and of the matter power spectrum, but also in the part of the code dealing with the galaxy number counts (including all relativistic effects). This part of the code was previously developed in flat space, in a branch called \classgal{}~\cite{DiDio:2013bqa}, which has been merged into the main \class{} public distribution since version {\sc v2.1.0}. The equations presented in this paper are included in the new release {\sc v2.5.0}.
We analyze to what extent geometrical (i.e., changes in the metric) and dynamical effects (i.e., changes in the background density parameters) due to curvature affect the angular power spectra.
We then investigate the effect of including relativistic terms on measurements of the curvature parameter by performing Fisher matrix forecasts on tomographic angle and redshift dependent power spectra.
We assume galaxy survey specifications consistent with a spectroscopic SKA-like \cite{Maartens:2015mra} experiment.

The paper is organized as follows.
In Section~\ref{sec:nonflat} we generalize the expression for the general relativistic galaxy number counts to non-flat geometries, and in Section~\ref{sec:cl} we introduce the observable we use to constrain curvature, i.e.~the redshift dependent angular power spectra.
Our results are discussed in Section~\ref{sec:measurements}.
In Section~\ref{sec:conclusions} we draw our conclusions.

\section{Galaxy number counts for non-flat universe}
\label{sec:nonflat}
In order to constrain the spatial curvature from LSS observations, we need to generalize LSS probes, in particular the galaxy number counts~\cite{Yoo:2009, Yoo:2010, Bonvin:2011bg,Challinor:2011bk}, to non-flat geometries for which we define the curvature $K$ by
\be
K = -H_0^2 \left(1 - \Omega_{\rm tot} \right) \;.
\ee
Being the galaxy number counts observables, they are gauge-invariant, so that we have the freedom to compute them in any arbitrary gauge. We choose to work in Newtonian gauge,
\be \label{metric}
d\tilde{s}^2 = a^2 \left( - \left( 1 + 2 \Psi \right) d\tau^2 + \left( 1 - 2 \Phi\right) \gamma_{ij} dx^i dx^j \right)
\ee
where:
\be
 \gamma_{ij} dx^i dx^j = \left[ dr^2 + \CHI^2 \left( r \right) \left( d\theta^2 + \sin^2 \theta d\varphi^2\right) \right]
\ee
and:
\bea
\CHI \left( r \right)= \left\{ \begin{array}{cc}
\frac{1}{\sqrt{K}}\sin \left( \sqrt{K}r  \right) & \text{for} \ K>0 \\
r & \text{for} \ K=0 \\
\frac{1}{\sqrt{\left| K \right| }} \sinh \left(\sqrt{\left| K \right| }r \right) & \text{for} \ K<0
\,.\end{array}
\right.   \eea
We consider only scalar perturbations,
i.e.~the Bardeen potentials $\Phi$ and $\Psi$, since to first order
vector, if generated, are diluted by the
expansion of the universe  and tensor perturbations are subleading and relevant only on very large scales. Nevertheless, they might be important at
second order.  We mainly follow the approach of~\cite{Bonvin:2011bg}
considering, however, the spatially curved the metric~(\ref{metric}).
We define the galaxy number counts in terms of the direction of observation $-\bn$ (where $\bn$ is the propagation direction of photons) and redshift $z$ as:
\be
\Delta \left( \bn, z \right) \equiv \frac{n_{\rm g}
  \left( \bn , z \right) - \langle n_{\rm g} \rangle \left( z \right)
}{\langle n_{\rm g} \rangle \left( z \right)} = \delta_z \left( \bn , z
\right) + \frac{\delta \nu \left( \bn , z \right)}{\bar \nu \left( z
  \right)}
\ee
where $\langle\ldots\rangle$ denotes the angular mean at fixed observed redshift $z$, $n_g(\bn, z)=dN/dz/d\Omega$ is the number density of sources per redshift and per solid angle and $\nu(\bn,z)=dV/dz/d\Omega$ is the volume density per redshift and per solid angle. We also introduced the density perturbation in redshift space
\be
\delta_z \left( \bn , z \right) =
\frac{\rho_{\rm g} \left( \bn , z \right) - \langle \rho_{\rm g} \rangle \left( z
  \right) }{\langle \rho_{\rm g} \rangle \left( z \right)} \;,
\ee
where $\rho_{\rm g}(\bn,z)=n_{\rm g}(\bn,z)/\nu(\bn,z)$ is the galaxy density per comoving volume, and the volume perturbation
\be
\delta \nu \left( \bn , z \right) = \nu\left(\bn , z \right) - \bar \nu \left( z \right)\;.
\ee
We remark that, not only the galaxy number counts $\Delta \left( \bn, z \right)$, but also the density perturbation in redshift space $\delta_z \left( \bn , z \right)$ and the volume perturbation $\delta \nu \left( \bn , z \right)/ \bar \nu \left( z \right)$ are gauge-invariant quantities. In the following we compute explicitly the contributions of these two terms.

\subsection{Density perturbation in redshift space}
As shown in~\cite{Bonvin:2011bg}, the redshift dependent density perturbation can be expanded to first order in perturbation theory as
\be \label{eq:redshift_den}
\delta_z  \left( \bn , z \right) = \frac{\delta \rho_{\rm g} \left( \bn , z \right)}{\bar \rho_{\rm g} \left( \bar z \right) } - \frac{d \bar \rho_{\rm g}}{d\bar z } \frac{\delta z \left( \bn , z \right)}{\bar \rho_{\rm g} \left( \bar z \right)}
\ee
where $ \bar z$ denotes the background redshift. In order to compute $\delta_z$ we need to derive, to first order, the redshift $z$ measured by an observer moving with a peculiar velocity $\bv_o$ and emitted by a source $\bv_s$.
We write the metric in eq.~(\ref{metric}) as $d \tilde s^2 = a^ 2 ds^2$ and use the fact that light-like geodesics are conformally invariant, i.e., $d \tilde s^2$ and $ds^2$ have the same light-like geodesics.
We choose the affine parameters in the two metrics $\tilde\lambda=a^2\lambda$, so that the photon geodesics are related by $\tilde n = n/a^2$.
Since we have $\tilde u = u/a$ for the 4-velocities, where $u=\left(1-\Psi, {\bf v}\right)$, the redshift is
\be
1+z = \frac{\left( \tilde n^\mu \tilde u_\mu \right)_s}{\left( \tilde n^\mu \tilde u_\mu \right)_o} = \frac{a_o}{a_s}\frac{\left( n^\mu u_\mu \right)_s}{\left( n^\mu u_\mu \right)_o} \, .
\ee
Being the spatial component $u^i$, i.e.~the peculiar velocity, already at first order, and normalizing the affine parameter $\lambda$ such that $\bar n^\mu = \left( 1 , \bn \right)$, we just need to compute $n^0$ by solving the geodesic equation,
\be \label{geodeq}
\frac{d \ \delta n^0 }{d\lambda} = \dot \Phi - \dot \Psi + 2 \partial_{r} \Psi =  \dot \Phi - \dot \Psi - 2 \bn \cdot \bnabla \Psi \, ,
\ee
where dots indicate derivatives with respect to conformal time $\tau$.
By applying the chain rule:
\be \label{chain_rule}
\frac{dA \left( \tau \left(\lambda \right) , \bx \left( \lambda \right) \right) }{d\lambda} = \frac{d A \left( \tau, \bx \left( \tau \right) \right) }{d\tau} = \dot A \left( \tau, \bx  \right) + \bn \cdot \bnabla A \left( \tau , \bx \right)
\ee
where $A \left( \tau \left(\lambda \right) , \bx \left( \lambda \right) \right) $ is an arbitrary first order quantity, we solve the geodesic equation~(\ref{geodeq})
\be
\delta n^0_o - \delta n^0 = \int_\tau^{\tau_o} \left( \dot \Phi + \dot \Psi \right) d\tau' -2 \Psi_o + 2 \Psi_s \, .
\ee
This yields to
\bea \label{redshift:1}
1 + z &=& \left( 1 + \bar z \right) \bigg( 1 + \Psi_o - \Psi_s + \bn \cdot \bv_o - \bn \cdot \bv_s
 - \int_\tau^{\tau_o} \left( \dot \Psi + \dot \Phi \right) d\tau' \bigg) \, .
\eea
It is not surprising that the curvature does not enter explicitly in the redshift expression. By adopting the Born approximation we consider photon traveling along unperturbed radial geodesic, while in the metric~(\ref{metric}) the curvature is confined in the generalized 2-sphere described by the 2-dimensional metric $ \left. \gamma_{ij}\right|_{r=1}$.
We neglect the quantities evaluated at the observer position that lead to a monopole and a dipole terms depending on the potential and the velocity, respectively. The former quantity is not observable, and the latter cannot be treated within cosmological perturbation theory.
Then, from eq.~(\ref{redshift:1}) we obtain
\be
\delta z = - \left( 1 + \bar z \right) \left( \Psi_s + \bn \cdot \bv_s + \int_\tau^{\tau_o} \left( \dot \Psi + \dot \Phi\right) d\tau' \right)
\, .
\ee
The Bianchi identity (or `energy' conservation equation) reads:
\be \label{cons_eq}
\frac{ d \bar \rho_{\rm g}}{dz} = 3 \frac{\bar \rho_{\rm g}}{1 + \bar z} \;.
\ee
This relation can be used to replace $\frac{1}{\bar\rho_{\rm g}}\frac{d\bar \rho_{\rm g}}{d\bar z} = \frac{1}{\bar n_{\rm g}}\frac{d\bar n_{\rm g}}{d\bar z}$, but it assumes $\bar \rho_{\rm g} a^3$ to be constant.
To be consistent with general models of galaxy formation we need to allow for a non-vanishing evolution term (also called evolution bias $b_e$ in the literature)
\be
f_\text{evo} \left( z \right) \equiv \frac{d \ln \left( a^3 \bar n_{\rm g} \right)}{\HH d\tau} = - \left( 1 + z \right) \frac{d}{dz} \ln \left( \frac{{\bar n}_{\rm g}}{\left( 1 + z \right)^3} \right) \, ,
\ee
so that
\be
\frac{1}{\bar n_{\rm g}}\frac{d\bar n_{\rm g}}{d\bar z}
= \frac{3}{1 + \bar z} - \frac{f_\text{evo}}{1 + \bar z} \;.
\ee
The redshift dependent density perturbation is
\bea
\delta_z \left( \bn , z \right)
&=& \delta_{\rm g} \left( \bn , z \right)
+ \left(3-f_\text{evo}\right)
\left[
  \Psi \left( \bn , z \right) + \bn \cdot \bv \left( \bn , z \right)
  +  \int_\tau^{\tau_o} \left( \dot \Psi + \dot \Phi\right) d\tau'
  \right]
\nonumber \\
&=&b D_{\rm m} \left( \bn , z \right) - (3-f_\text{evo}) \HH v
+ \left(3-f_\text{evo}\right)
\Big[ \Psi  \left( \bn , z \right)+  \bn \cdot \bv \left( \bn , z \right)
  \nonumber \\
  &&
  +  \int_\tau^{\tau_o} \left( \dot \Psi + \dot \Phi\right) d\tau'
  \Big]\;. \label{den:red}
\eea
We have introduced the galaxy density in Newtonian (or longitudinal) gauge $\delta_{\rm g}$, and the gauge invariant variable $D_{\rm m}$ that coincides with the dark matter density perturbation in the comoving gauge (see~\cite{DurrerBook} for more details).
At first order in perturbation theory, $D_{\rm m}$ also corresponds to the density perturbation in synchronous gauge comoving with dark matter.
We assume that it is related to the comoving gauge galaxy density perturbation $D_{\rm g}$ by a linear galaxy bias $D_{\rm g}=bD_{\rm m}$, which can be function of time and of the scale.
This galaxy bias prescription is justified from the fact that on large enough scales, both galaxies and dark matter follow the same velocity field as they experience the same gravitational acceleration \cite{Baldauf:2011bh,Jeong:2011as}.
Hence, we further assume no velocity bias, so that by $v$ we indicate at the same time the (gauge invariant) matter and the galaxy velocity potential in the Newtonian gauge, defined through $\bv = -\nabla v$.
The gauge transformation relating the galaxy density perturbations in the longitudinal and synchronous gauges, taking into account also evolution bias \cite{Challinor:2011bk}, is then:
\be
\label{eq:gauge_gal}
\delta_{\rm g} = D_{\rm g} - \left( 3 - f_\text{evo}  \right)\HH v \;.
\ee

\subsection{Volume perturbation}
Now, we compute the contribution to the galaxy number counts induced by the volume perturbation. We start considering an infinitesimal volume element around the source defined by
\be
dV = \sqrt{-g} \epsilon_{\mu \si \alpha \beta} u^\mu dx^\si dx^\alpha dx^\beta \, .
\ee
Since we are interested in expressing all the perturbations in terms of observable quantities, we expand the angles at source around the observed angles at the observer,
\be
dV = \nu \left( z , \theta_o, \varphi_0 \right) d z d \theta_o d\varphi_o
\ee
where
\be \label{vol:den}
\nu \left( z , \theta_o, \varphi_0 \right) = \sqrt{-g} \epsilon_{\mu \si \alpha \beta} u^\mu \frac{\partial x^\si}{\partial z} \frac{\partial x^\alpha}{\partial \theta_s} \frac{\partial x^\beta}{\partial \varphi_s} \left| \frac{\partial \left( \theta_s , \varphi_s \right) }{\partial \left( \theta_o , \varphi_o \right)} \right|
\ee
and $\left| \frac{\partial \left( \theta_s , \varphi_s \right) }{\partial \left( \theta_o , \varphi_o \right)} \right|$ is the determinant of the Jacobian of the coordinate transformation from the angles at the source $\left( \theta_s , \varphi_s \right)$ to the angles at the observer $\left( \theta_o, \varphi_o \right)$. To first order in perturbation theory it becomes
\be
\left| \frac{\partial \left( \theta_s , \varphi_s \right) }{\partial \left( \theta_o , \varphi_o \right)} \right| = 1 + \frac{\partial \delta \theta}{\partial \theta} + \frac{\partial \delta \varphi}{\partial \varphi} \, .
\ee
Then, by writing explicitly~(\ref{vol:den}), we find
\be
\label{vol:den:2}
\nu = a^4 \left( 1 + \Psi - 3 \Phi \right) \left[ \CHI^2(r) \sin \theta_s \right]
 \left( \frac{1}{a} \left( 1 - \Psi \right)  \frac{dr}{dz} - \frac{1}{a} v^{r} \frac{d\tau}{dz} \right)
 \left( 1 + \frac{\partial \delta \theta}{\partial \theta} + \frac{\partial \delta \varphi}{\partial \varphi} \right) \, .
\ee
We need to compute how the radial coordinate $r$ changes along the light geodesic,
\bea
\frac{dr}{dz} &=& \frac{d}{d \bar z} \left( \bar r + \delta r \right) \frac{d }{dz} \left( z - \delta z \right)
=  \frac{d\tau}{d\bar z } \left( \frac{d \bar r }{d \tau} + \frac{d \delta r }{d \lambda} - \frac{d \bar r }{d \bar z } \frac{d \delta z }{d \lambda} \right)
\eea
and expand around the background the following term:
\be
\CHI^2(r)\sin \theta_s =\CHI^2\left( \bar r+ \delta r \right) \sin \left( \theta_o + \delta \theta \right)
= \CHI^2\left( \bar r \right) \sin \theta_o  \left( 1 + 2 \frac{\CHI' \left( \bar r \right)}{\CHI \left( \bar r \right)} \delta r + \cot \theta_o \delta \theta \right) \, .
\ee
With these two expressions, we can rewrite~(\ref{vol:den:2}) as
\be
\nu = \frac{a^4}{\HH} \CHI^2  \sin \theta
\left[ 1 - 3 \Phi - \bn \cdot \bv_s + \left( \cot \theta + \frac{\partial}{\partial \theta} \right) \delta \theta + \frac{\partial \delta \varphi}{ \partial \varphi}
+ \left(
2 \frac{\CHI'}{\CHI}- \frac{d}{d \lambda} \right) \delta r + \frac{a}{\HH} \frac{d \delta z }{d \lambda} \right] \, ,
\ee
where all quantities are evaluated at the source background position.
Then, we expand $\bar \nu \left( z \right)$ around the background redshift $\bar z$ obtaining
\bea
\bar \nu \left( z \right) &=& \bar \nu \left( \bar z \right) + \frac{d \bar \nu }{d \bar z } \delta z
= \bar \nu \left( \bar z \right) \left[ 1 + \frac{\delta z }{1 + \bar z } \left( \frac{2}{\HH}\frac{\CHI'}{\CHI} - 4 + \frac{\dot \HH }{\HH^2} \right)\right] \, .
\eea
This leads to the volume perturbation,
\bea
\frac{\delta \nu}{\bar \nu } \left( \bn , z \right) &=& - 3 \Phi - \bn \cdot \bv + \left( \cot \theta + \frac{\partial}{\partial \theta} \right) \delta \theta + \frac{\partial \delta \varphi}{\partial \varphi}
 + \left( 2\frac{\CHI'}{\CHI}- \frac{d}{d\lambda} \right) \delta r
\nonumber \\
&&
+ \left( 4 - \frac{2}{\HH}\frac{\CHI'}{\CHI}- \frac{\dot \HH}{\HH^2} \right) \frac{\delta z}{1+ \bar z }
+ \frac{1}{\HH(1+ \bar z)} \frac{d \delta z}{d\lambda} \, .
\eea
To completely determine the volume perturbation we need to solve the geodesic equations for $\delta r$, $\delta \theta$, $\delta \varphi$.
We start from the radial coordinate,
\bea
\frac{dr}{d\tau} &=& \frac{dr}{d\lambda} \frac{d\lambda}{d\tau} = \frac{-1 + \delta n^{r}}{1 + \delta n^0}= -1 +\delta n^{r} + \delta n^0
= -1 + \frac{ d \delta r}{d\tau}  \,.
\eea
From the geodesic equation
\be
\frac{d  \delta n^{r}}{d\lambda} =\partial_{r} \Phi - \partial_{r} \Psi - 2 \dot \Phi
\ee
and eq.~(\ref{geodeq}) we obtain
\bea
\frac{d^2 \delta r}{d\tau^2} &=& \partial_{r} \Phi +\partial_{r} \Psi -  \dot \Phi   - \dot \Psi
= - \frac{d}{d\lambda} \left( \Psi + \Phi \right)
\eea
from which it follows
\be
\delta r = \int_\tau^{\tau_o} \left( \Psi + \Phi \right) d\tau'
\ee
where we have again neglected the contributions at the observer position.
So, the radial contribution to the volume perturbation is
\bea
&&\left( 2\frac{\CHI'}{\CHI} - \frac{d}{d \lambda} \right) \delta r
= 2 \frac{\CHI'}{\CHI} \int_\tau^{\tau_o} \left( \Psi + \Phi \right) d\tau' + \Psi + \Phi \, .
\eea
After the radial coordinate, we look at the angles $\theta$ and $\varphi$. We have
\bea \label{vol:33}
&&\frac{d\theta}{d\tau} = \frac{d\theta}{d \lambda} \frac{d \lambda}{d\theta}= \frac{\delta n^\theta}{1 + \delta n^0}= \delta n^\theta = \frac{d \delta \theta}{d\tau} \, , \\
&& \frac{d\varphi}{d\tau} = \frac{d\varphi}{d \lambda} \frac{d \lambda}{d\varphi}= \frac{\delta n^\varphi}{1 + \delta n^0}= \delta n^\varphi  = \frac{d \delta \varphi}{d\tau}\, . \label{vol:34}
\eea
The geodesic equations lead to
\bea
&&\frac{d \delta n^\theta}{d\lambda} - 2 \left(  \frac{d}{dr} \ln \CHI \right) \delta n^\theta = -\frac{1}{\CHI^2} \partial_\theta \left( \Psi + \Phi \right) \nonumber \\
&\Rightarrow& \frac{d}{d\lambda} \left( \delta n^\theta \CHI^2 \right) = - \partial_\theta \left( \Psi + \Phi \right)
\eea
which is solved by
\be
\delta n^\theta = \frac{1}{\CHI^2} \int_\tau^{\tau_o} \partial_\theta \left( \Psi + \Phi \right) d\tau' \, .
\ee
Analogously, we have
\bea
&&\frac{d \delta n^\varphi}{d\lambda} - 2 \left( \frac{d}{dr} \ln \CHI \right) \delta n^\varphi = - \frac{1}{\CHI^2 \sin^2 \theta} \partial_\varphi \left( \Psi + \Phi \right) \nonumber \\
&\Rightarrow& \frac{d}{d\lambda} \left( \delta n^\varphi \CHI^2 \right) = -\frac{1}{\sin^2\theta} \partial_\varphi \left( \Psi + \Phi \right)
\eea
which is solved by
\be
\delta n^\varphi = \frac{1}{\CHI^2 \sin^2\theta} \int_\tau^{\tau_o} \partial_\varphi \left( \Psi + \Phi \right) d\tau' \, .
\ee
From eqs.~(\ref{vol:33}, \ref{vol:34}) it follows
\bea
\delta \theta &=& - \int_\tau^{\tau_o} \frac{d \tau'}{\CHI^2} \int_{\tau'}^{\tau_o} d\tau'' \partial_\theta  \left( \Psi + \Phi \right) \, , \\
\delta \varphi &=& - \int_\tau^{\tau_o} \frac{d \tau'}{\CHI^2 \sin^2\theta} \int_{\tau'}^{\tau_o} d\tau'' \partial_\varphi \left(  \Psi + \Phi \right) \, .
\eea
Hence, the angular contribution to the volume perturbation is given by
\bea
\left( \cot \theta + \frac{\partial}{\partial \theta} \right) \delta \theta + \frac{\partial \delta \varphi}{\partial \varphi}
&=& - \int_\tau^{\tau_o} \frac{d\tau'}{\CHI^2 } \int_{\tau'}^{\tau_o} d\tau'' \left[ \cot \theta \frac{\partial}{\partial \theta} + \frac{\partial^2}{\partial \theta^2}
  + \frac{1}{\sin^2\theta} \frac{\partial^2}{\partial\phi^2} \right] \left( \Phi + \Psi \right)
 \nonumber \\
 &=& - \int_\tau^{\tau_o} \frac{d\tau'}{\CHI^2} \int_{\tau'}^{\tau_o} d\tau'' \Delta_\Omega \left( \Psi + \Phi \right) \, ,
\eea
where $\Delta_\Omega = \left( {\rm cot} \theta \partial_\theta + \partial_\theta^2+\partial_\varphi^2/{\rm sin}^2 \theta \right) $ is the angular part of the Laplacian.
This expression can be further simplified by partial integration and using
\be
\frac{d}{dr} \frac{\CHI'}{\CHI}= - \frac{1}{\CHI^2}\, .
\ee
As expected, we find that the angular contribution to the volume perturbation describes the lensing effect in the generalized form
\be
-  \int_\tau^{\tau_o} \left(\frac{\CHI' \left( r' \right)}{\CHI\left( r \right)} -  \frac{\CHI' \left( r \right)}{\CHI \left( r\right)}  \right) \Delta_\Omega \left( \Psi + \Phi \right) d\tau' \, ,
\ee
where $r'$ denotes $\tau_o - \tau'$.
By using the simple relation
\be
\CHI\left( r_s - r \right) = \CHI\left( r_s \right) \CHI' \left(r \right) - \CHI'\left( r_s \right) \CHI \left( r \right) \;,
\ee
we obtain the lensing term in a more common form,
\be
-  \int_\tau^{\tau_o} \frac{\CHI\left( r - r' \right)}{\CHI \left( r \right) \CHI \left( r' \right)} \Delta_\Omega \left( \Psi + \Phi \right) d\tau' \, .
\ee
To conclude, by plugging together all the contributions, we find the total volume perturbation
\bea
\frac{\delta \nu}{\bar \nu} &=& -2 \left( \Psi + \Phi \right)
+ \frac{\dot \Phi}{\HH} + \frac{1}{\HH} \partial_{r} \bn \cdot \bv + \bn \cdot \bv
-  \int_\tau^{\tau_o} \frac{\CHI \left( r - r' \right)}{\CHI \left( r \right) \CHI \left( r' \right)} \Delta_\Omega \left( \Psi + \Phi \right) d\tau'
\nonumber \\
&&
+2\frac{\CHI'}{\CHI} \int_\tau^{\tau_o} \left( \Psi + \Phi \right) d\tau' - 4 \bn \cdot \bv - 3 \int_\tau^{\tau_o} \left( \dot \Psi + \dot \Phi \right) d \tau'
\nonumber \\
&&
+ \left( \frac{2}{\HH}\frac{\CHI'}{\CHI}  + \frac{\dot \HH}{\HH^2} \right) \left( \Psi + \bn \cdot \bv + \int_\tau^{\tau_o} \left( \dot \Psi + \dot \Phi \right) d\tau' \right)
 \;, \label{vol:per}
\eea
where we have assumed that galaxies moves along geodesics,
\be
\bn \cdot \dot\bv + \HH \bn \cdot \bv - \partial_{r}\Psi=0 \;.
\ee

By summing up the redshift density perturbation~(\ref{den:red}) with the volume perturbation~(\ref{vol:per}) we find the galaxy number counts
\bea \label{eq:NC}
\Delta \left( \bn , z \right) &=& b D_{\rm m} - 2 \Phi + \Psi + \frac{1}{\HH} \left( \dot \Phi +  \partial_{r} \bn \cdot \bv \right)
- \int_\tau^{\tau_o}\frac{\CHI\left( r - r' \right)}{\CHI \left( r \right) \CHI \left( r' \right)} \Delta_\Omega \left( \Psi + \Phi \right) d\tau'
\nonumber \\
&&
+ \left( \frac{2}{\HH} \frac{\CHI'}{\CHI} + \frac{\dot \HH}{\HH^2} - f_\text{evo} \right) \left( \Psi + \bn \cdot \bv + \int_\tau^{\tau_o} \left( \dot \Psi + \dot \Phi \right) d\tau' \right)
\nonumber \\
&& + 2\frac{\CHI'}{\CHI} \int_\tau^{\tau_o}  \left( \Psi + \Phi \right) d\tau' - (3 - f_\text{evo}) \HH v \, .
\eea
In the flat limit one needs just to replace $\CHI'/\CHI$ with $1/r$ to recover the expression computed in~\cite{Bonvin:2011bg}.

\subsection{Magnification bias}

Another effect to take into account is induced by the flux limited properties of galaxy surveys. Up to now we have implicitly assumed that galaxy surveys are volume limited, i.e.~we observe all the galaxy up to a given redshift. Flux limited galaxy survey may observer fainter galaxies if their signal is magnified along the line of sight, and viceversa \cite{Bartelmann:1999yn}. Hence, this effect might be degenerate with the lensing (magnification) signal and we need to carefully introduce it, see e.g.~\cite{Challinor:2011bk, DiDio:2013bqa} for $K=0$. Including this effect, so considering that the observed galaxy number counts depends also on the luminosity, at a given fixed flux $F$, it becomes
\be
\Delta \left( \bn , z , F\right) = \Delta \left( \bn , z \right) + \left. \frac{\partial \ln \bar n_{\rm g}}{\partial \ln L_s} \right|_{\bar L_s} \frac{\delta L_s}{\bar L_s} \, ,
\ee
where $\bar L_s$ is the background luminosity corresponding to the flux $F$.
The fractional fluctuation in the luminosity at fixed flux is given by twice the fractional fluctuation in the luminosity distance, i.e.~$
\delta L_s/\bar{L}_s = 2 \delta D_L/\bar{D}_L$; we compute $\delta D_L$ to first order for non-flat geometries, i.e.~for the metric~(\ref{metric}), in Appendix~\ref{sec:distance}. Hence we obtain
\bea
\Delta \left( \bn , z ,m_* \right) &=& b D_{m}  + \frac{1}{\HH} \partial_r \left( \bn \cdot \bv \right)  - \frac{2 -5s  }{2} \int_\tau^{\tau_o}  \frac{\CHI\left( r - \tilde r \right)}{\CHI \left( r \right) \CHI \left( \tilde r \right)} \Delta_\Omega \left( \Psi + \Phi \right) d\tilde\tau
\nonumber \\
&&
 + \left( 5 s +\frac{2-5s}{\HH} \frac{\CHI'}{\CHI} + \frac{\dot \HH}{\HH^2} - f_\text{evo} \right) \left( \Psi + \bn \cdot \bv + \int_\tau^{\tau_o} \left( \dot \Psi + \dot \Phi \right) d\tilde\tau \right)
\nonumber \\
&&
+ \left( 5s- 2\right) \Phi + \Psi + \frac{1}{\HH} \dot \Phi + (f_\text{evo}-3) \HH v
\nonumber \\
&&
+ \left( 2 -5s \right) \frac{\CHI'}{\CHI} \int_\tau^{\tau_o}  \left( \Psi + \Phi \right) d\tilde\tau
 \; .  \label{eq:NC_obs}
\eea
We defined the magnification bias (see~\cite{DiDio:2013bqa} for more details)
\bea
\frac{\partial \ln \bar n_{\rm g} \left( z , L > \bar L_* \right)}{\partial \ln L_*} = - \frac{5}{2} s \left( z, m_* \right) \;,
\eea
where
\be
\bar n_{\rm g} \left( z , L > \bar L_* \right) = \int_{F_*}^\infty \bar n_{\rm g} \left( z , \ln F\right) d \ln F
\ee
denotes the background number density of galaxies with luminosity larger than $L_*$ (or equivalently a magnitude smaller than $m_*$ or flux larger than $F_*$).
The magnification bias $s$ is also often replaced by $Q=5s/2$ in the literature.

\section{The angular power spectrum}
\label{sec:cl}
The power spectrum cross-correlating two scalars $X$ and $Y$ is
\be
C_{\ell}^{XY}(z,z')=\left< a^X_{\ell m}(z) a^{Y\,^*}_{\ell m}(z') \right> \;,
\ee
where the star denotes complex conjugation and $X=\sum a_{\ell m}(z)Y_{\ell m}(\bn)$, where $Y_{\ell m}(\bn)$ denotes the spherical harmonic functions~\cite{abramowitz}.
Given the transfer function $\Delta_{\ell}^{X}(q,z)$ the angular power spectra for an open universe $K\leq0$ reads \cite{Hu:1997mn,DurrerBook,Lesgourgues:2013bra}:
\begin{equation} \label{eq:Cl_def}
C_{\ell}^{XY}(z,z') = 4\pi \int \frac{dk}{k} \Delta_{\ell}^{X}(q,z) \Delta_{\ell}^{Y}(q,z') \mathcal{P}_{\mathcal{R}}(k) \;,
\end{equation}
where the generalized wavenumber $q=\sqrt{k^2+K}$ is seen as a function of $k$ (wavenumber in the flat limit), and $\mathcal{P}_{\mathcal{R}}(k)$ is the primordial spectrum of scalar curvature perturbations.
  We remark that a power-law power spectrum is uniquely determined in terms of wavelength $k$, generalized wavenumber $q$, which is an eigenvalue of the Helmholtz equation~(\ref{helmotz_eq}), and volume~(see e.g.~\cite{Kamionkowski:1994qf}) only for spatially flat space-time. For open and close universe they differ for modes larger than the curvature radius. Our choice, consistently with~\cite{Lesgourgues:2013bra}, is to consider a power-law in the wavelength $k$, namely $\mathcal{P}_{\mathcal{R}}(k) \propto k^n$. Nevertheless, since the ratio between the curvature and the Hubble radius is determined by $\sqrt{\Omega_K}$, a curvature density parameter of order $\Omega_K \sim \mathcal{O} \left(10^{-2} \right)$ leads to observational differences only at scale of order $\mathcal{O} \left( 10 \right) $ Hubble radius. Hence our choice does not affect our results, and generically late time observables.
We also introduce a rescaled generalized wavenumber\footnote{We remark that in literature the rescaled generalized wavenumber is often denoted with $\nu$. We prefer to use $\kappa$, since in this work $\nu$ refers to the volume density defined through eq.~(\ref{vol:den}).} $\kappa=q/\sqrt{|K|}$.
For a closed universe $K>0$ the integral in eq.~(\ref{eq:Cl_def}) is replaced by a sum over discrete $k_n=\sqrt{(n^2-1)K}$ related to generalized wave numbers by $\kappa_n=q_n/\sqrt{K}=n$ with $n=3,4,5,\ldots$

We compute the angular power spectra using the \class{} code \cite{Lesgourgues:2011re,Blas:2011rf,Lesgourgues:2013bra}.
We obtain the transfer functions $\Delta_{\ell}^{X}(q,z)$ by expanding eq.~(\ref{eq:NC_obs}) over the basis $Q_{\bk}(\bx)$ given by the eigenfunctions of the Laplacian
\begin{equation} \label{helmotz_eq}
\left(\Delta_K+k^2\right)Q_{\bk}(\bx)=0\;.
\end{equation}
In the flat limit $Q_{\bk}(\bx)=\exp(i{\bf k}\cdot{\bf x})/(2\pi)^{3/2}$ and its expansion in spherical harmonics is written in terms of spherical Bessel functions $j_{\ell}[k(\tau_0-\tau)]$.
Following \cite{Abbott:1986ct, Kamionkowski:1994qf, Hu:1997mn, DurrerBook, Lesgourgues:2013bra}, the generalization to a non-flat spacetime involves hyperspherical Bessel functions $\Phi_{\ell}^{\kappa}(\chi)$, where we use the rescaled radial coordinate $\chi=\sqrt{|K|}(\tau_0-\tau)$ such that in the flat limit $\Phi_{\ell}^{\kappa}(\chi) \to j_{\ell}(\kappa\chi)=j_{\ell}[k(\tau_0-\tau)]$.
The number count transfer functions obtained from eq.~(\ref{eq:NC_obs}) are shown in Appendix~\ref{s:transfer}, while in Appendix~\ref{s:harmonic} we also give more details about scalar harmonic functions $Q_{\bk}(\bx)$.

\section{Constraining the curvature parameter}
\label{sec:measurements}

In this section we consider a mock galaxy survey loosely modeled after the SKA HI galaxy survey~\cite{SKA:Abdalla} (see also e.g.~\cite{Camera:2014bwa, Santos:2015hra} for a discussion on possible SKA surveys).
In Appendix~\ref{sec:forecast} we show the specifications used in our calculations and describe our Fisher matrix formalism.
We investigate how effects which are usually neglected in galaxy clustering analysis affect constraints on the curvature parameter.

\subsection{Standard Newtonian and full relativistic spectra}
\label{sec:Cl_plots}

In the following we investigate how considering a fully relativistic tomographic analysis instead of a traditional Newtonian one affects the angular power spectra for non-flat universes. We define the ``standard Newtonian'' analysis one which  includes in eq.~(\ref{eq:Cl_def}) only the density perturbation and the Kaiser redshift-space distortions (however we do not apply the plane-parallel approximation):
\be \label{eq:D_newt}
\Delta_{\ell}^{\rm Newt} = \Delta_{\ell}^{\mathrm{Den}} + \Delta_{\ell}^{\mathrm{Rsd}}\;.
\ee
In the ``fully relativistic'' analysis we consider all contributions, including lensing convergence \cite{Bartelmann:1999yn, Matsubara:2000pr, LoVerde:2007ke},\footnote{While the lensing potential is a well known and constrained term from galaxy shear studies, in the context of galaxy clustering it is usually neglected, except for specific configurations~\cite{Montanari:2015rga}.} additional velocity terms (that we call Doppler, D1 and D2), and effects depending directly on the gravitational potentials (G1 to G5):
\be \label{eq:D_rel}
\Delta_{\ell} = \Delta_{\ell}^{\mathrm{Den}} + \Delta_{\ell}^{\mathrm{Rsd}} + \Delta_{\ell}^{\mathrm{Len}} + \sum_{a=1}^2 \Delta_{\ell}^{\mathrm{D}a} + \sum_{a=1}^5 \Delta_{\ell}^{\mathrm{G}a} \;,
\ee
where all the terms are defined in Appendix~\ref{s:transfer}, eq.~(\ref{eq:delta_terms}).
We always consider redshift bin cross-correlations, both for Newtonian and relativistic terms. As it has been shown~\cite{Asorey:2012rd,DiDio:2013sea}, this is needed to recover the 3-dimensional information in the galaxy catalog. We stress that in our conservative configuration, i.e.~5 redshift bins, cross-correlations add constraining power mostly through the non-local relativistic terms (cosmic magnification, ISW and time delay effects, see~\cite{Raccanelli:radial}).

\begin{figure}[t!]
  \center{
    \includegraphics[width=0.48\linewidth]{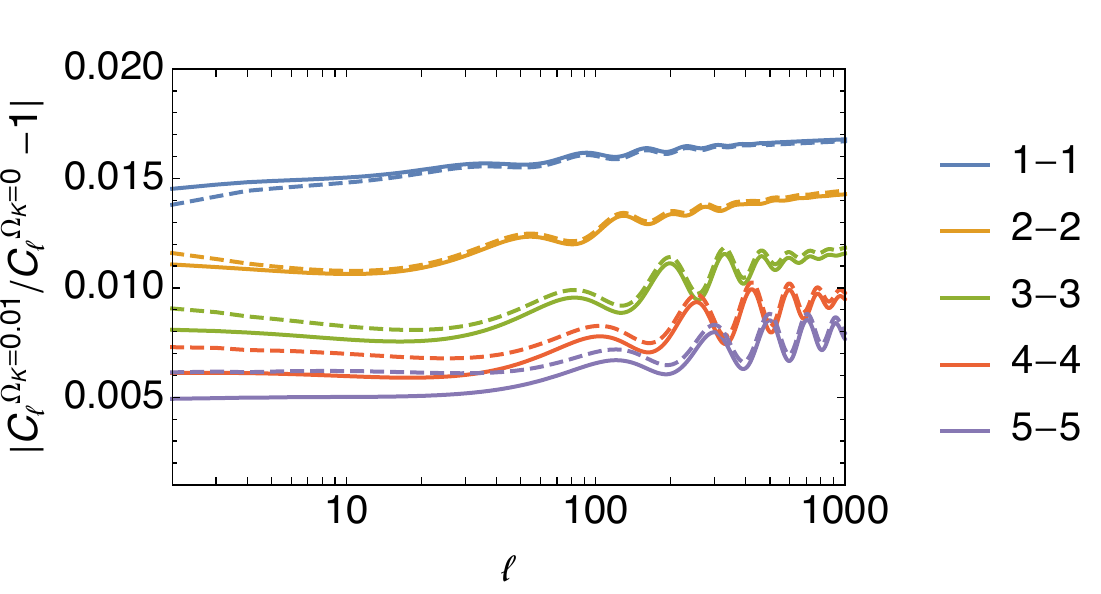}
    \quad
    \includegraphics[width=0.48\linewidth]{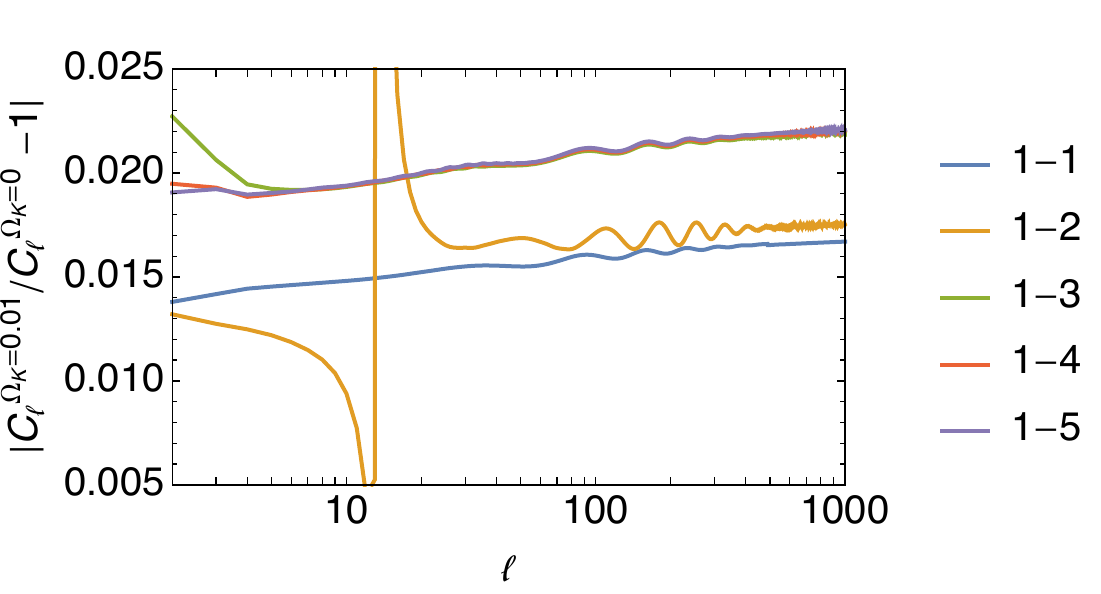}
  }
  \caption{Relative variation of angular power spectra when including a spatial curvature $\Omega_K=0.01$, assuming top-hat redshift bins. Left panel shows z-bin auto-correlations while right panel z-bin cross-correlations. In the left panel we show the ratio by including all the relativistic effects (solid lines) and considering only Newtonian terms (dashed lines).  }
  \label{fig:Cls_kai}
\end{figure}

In figure~\ref{fig:Cls_kai} we plot the relative variation of the angular power spectra when adding a non-vanishing curvature parameter $\Omega_K=0.01$. We plot
$$
\frac{C_\ell^{\Om_K=0.01}(z_i,z_j) -C_\ell^{\Om_K=0}(z_i,z_j)}{C_\ell^{\Om_K=0}(z_i,z_j)}\,.
$$
We fix $\Omega_m=0.3$ and we vary $\Omega_\Lambda$ according to eq.~(\ref{eq:fried_constr}). We choose 5 redshift bins as specified in Appendix~\ref{sec:forecast}, fig.~\ref{fig:SKA_spec}.
For the redshift bin auto-correlations we show the effect of full power spectra, i.e.~including both Newtonian and relativistic contributions, and we compare them with just the Newtonian terms. For cross-correlation we only plot the full power spectra, since the Newtonian one would be dominated by numerical noise.
Not surprisingly, the first bin is most strongly affected by curvature as it becomes more relevant at low redshift. For all except the first bin, including relativistic effects enhances the difference between the curved and flat spectra.
Note also the change of sign in the 1-2 cross correlation at $\ell=13$ which comes from the zero-crossing of the lensing term $\propto 2-5s(z)$ which happens in the second bin. The cross-correlation signal is dominated by the lensing terms as already discussed in Ref.~\cite{Montanari:2015rga}.
Obviously a non-vanishing curvature changes the angular diameter distance and shifts the BAOs in angular scale. This effect causes the wiggles in the left panel of figure~\ref{fig:Cls_kai}. Also the change in $\Omega_\Lambda$, which varies together with $\Omega_K$ according to eq.~(\ref{eq:fried_constr}) leads to a difference in the angular diameter distance.
We study the impact of the differences, induced by relativistic terms (here dominated by the lensing term), on the precision of measurements of the curvature parameter and its best-fit estimate in the next sections.

\begin{figure}[t!]
  \center{
    \includegraphics[width=0.48\linewidth]{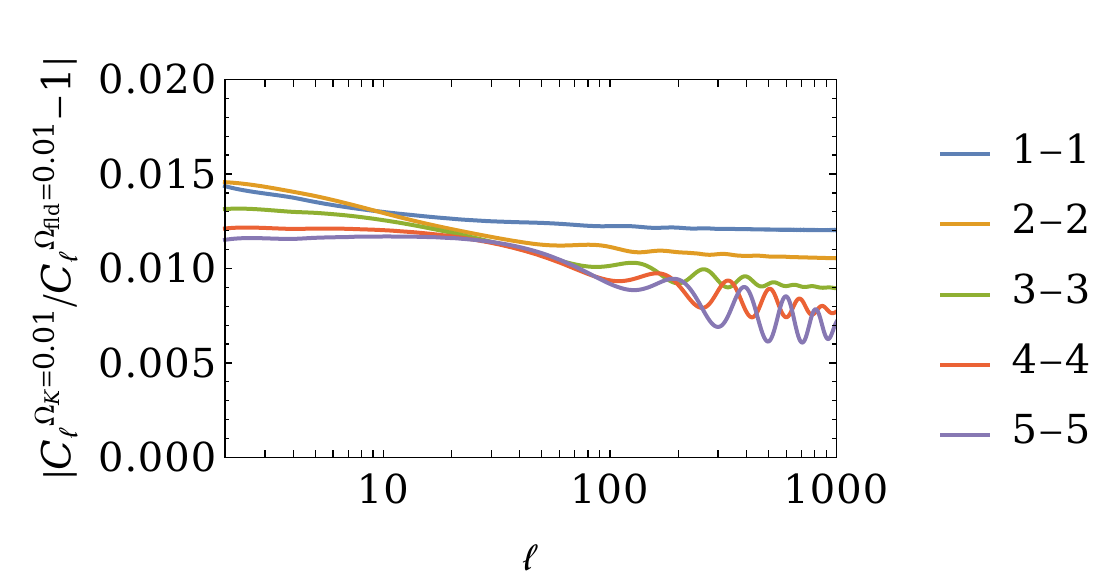}
    \quad
    \includegraphics[width=0.48\linewidth]{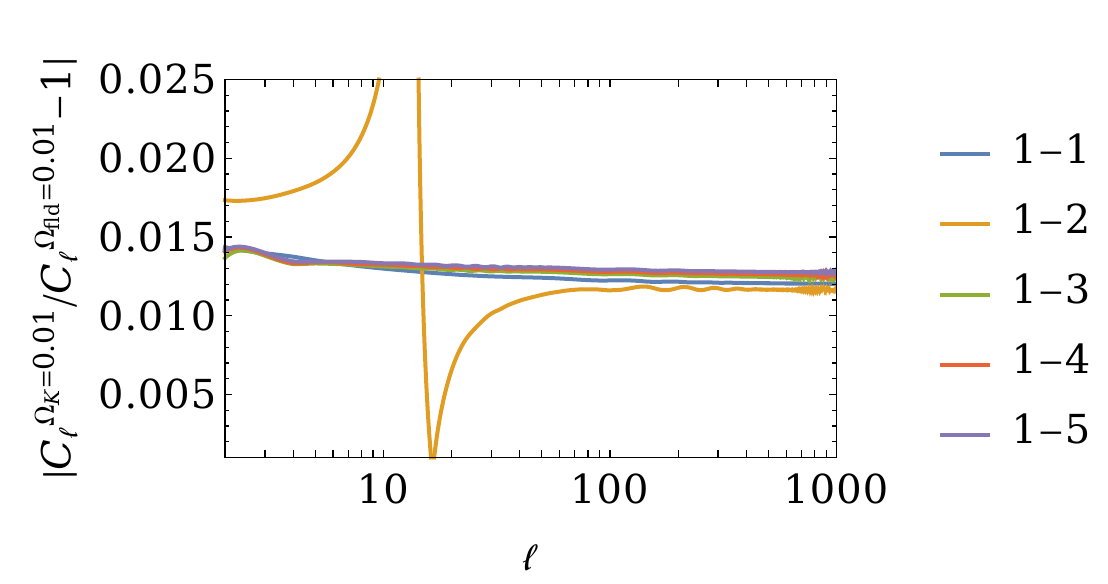}
  }
  \caption{
    Importance of geometrical contributions: the relative difference of the relativistic angular power spectra for the o$\Lambda$CDM and the f$\Lambda$CDM model is plotted as a function of $\ell$. The dynamics of the background evolution is the same in both cases, but geometrical changes due to curvature are not present in the fluid model.
    The left panel shows z-bin auto-correlation while the right panel z-bin cross-correlation. Bin correlations are showed in the legends.
    The spike in the bins 1-2 cross-correlation happens when the spectrum at the denominator passes through zero.
  }
  \label{fig:dyn_geom}
\end{figure}

\begin{figure}[t!]
  \center{
    \includegraphics[width=0.48\linewidth]{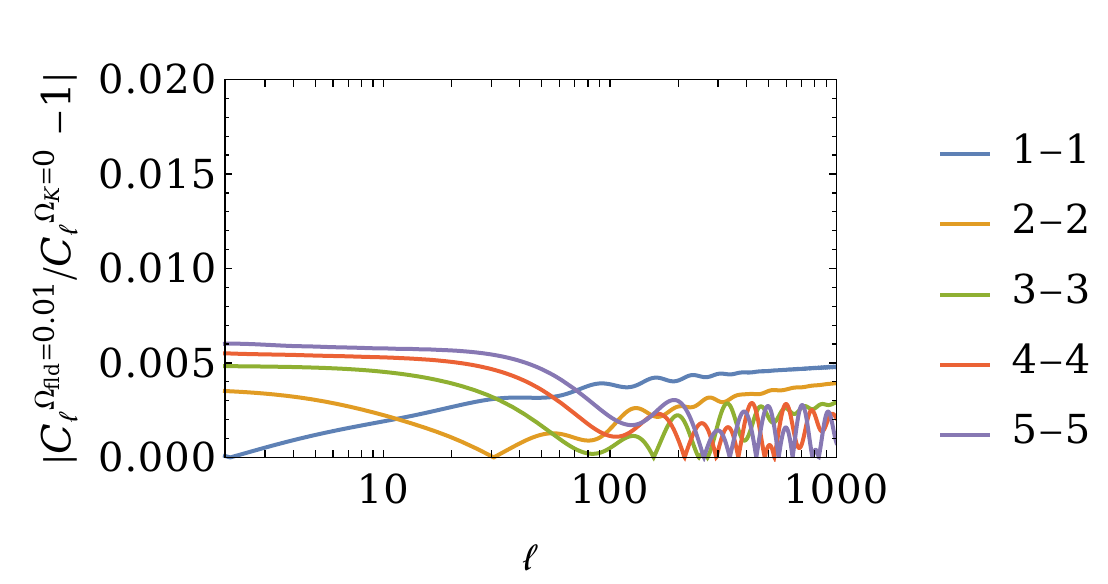}
    \quad
    \includegraphics[width=0.48\linewidth]{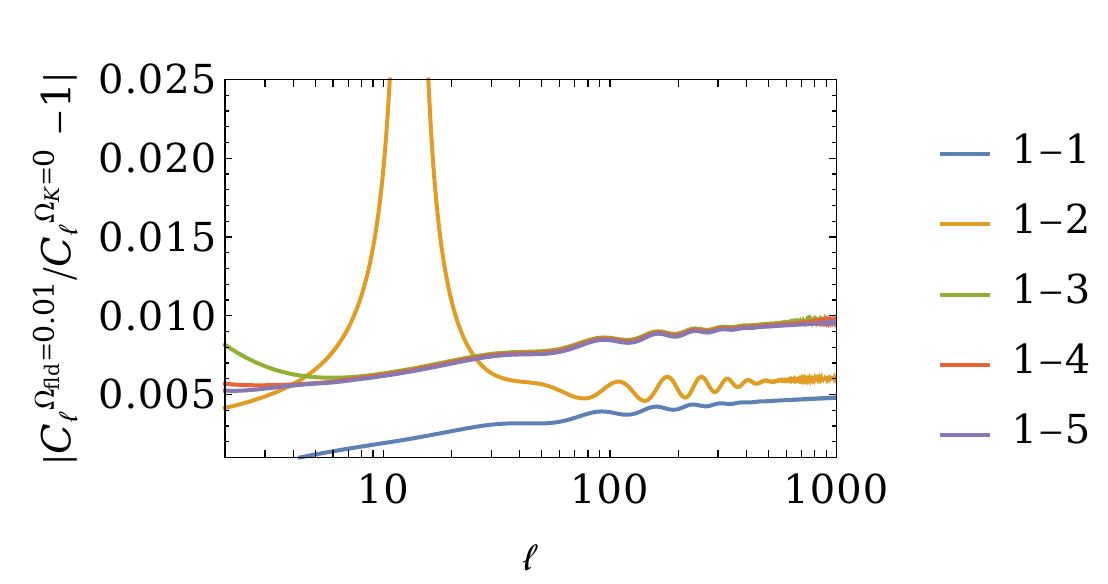}
  }
  \caption{
    Importance of dynamical contributions: the relative variation of the relativistic angular power spectra for the f$\Lambda$CDM model is plotted against the $\Lambda$CDM model.
    While geometrical terms due to curvature are absent in both cases, the background evolution changes.
    The curve legend follows figure~\ref{fig:dyn_geom}.
    While larger oscillations reflect differences in the acoustic peaks location, smallest scales oscillation are due to numerical precision and interpolation.
  }
  \label{fig:dyn_geom_2}
\end{figure}

To disentangle geometrical (changes in the metric) and dynamical (changes in the background density parameters and therefore in the Hubble parameter, $H(z)$) effects due to curvature, we introduce a fictitious non-adiabatic fluid with roughly constant equation of state\footnote{For an adiabatic perfect fluid we have $\dot{w}=3\HH(1+w)(w-c_s^2)$, that requires either $w=-1$ or $w=c_s^2$ in order to have a constant $w$ over time. For more details about the implementation of the non-adiabatic fluid with a generic nearly constant equation of state, see \cite{Ballesteros:2010ks}.} $w_{\rm fld}\approx-1/3$, whose density parameter scales in the same way as the one from curvature, while keeping a flat metric.
Since we are not interested in the dynamics of the non-adiabatic fluid itself, but only in understanding by how much power spectra are affected by changes in the density parameters, we neglect its perturbations\footnote{Note, however, that in the public version of \class{}, fluid perturbations are consistently taken into account.} (in particular, our results are independent of the fluid sound speed $c_s$).
This artifact allows us to take into account changes to the Hubble parameter $H(z)$ due to different density parameters, still keeping a flat metric.
\begin{table}[t!]
  \begin{center}
    \begin{tabular}{|r|c|c|c|c|}
      \hline
      & $\Omega_{\Lambda}$  & $\Omega_K$ & $\Omega_{\rm fld}$ & $w_{\rm fld}$\\
      \hline
      $\Lambda$CDM      & 0.7 & 0 & 0 & \\
      o$\Lambda$CDM & 0.69 & 0.01 & 0 & \\
      f$\Lambda$CDM & 0.69 & 0 & 0.01 & -1/3\\
      \hline
    \end{tabular}
    \caption{
      Cosmologies used to disentangle dynamical and geometrical contributions from the curvature parameter: standard flat model $\Lambda$CDM, open model o$\Lambda$CDM, and flat model f$\Lambda$CDM including a non-adiabatic fluid mimicking the dynamical effects of curvature. In all the cases we fix $\Omega_{\rm m}=0.3$.
    }
    \label{tab:cosmo}
  \end{center}
\end{table}
In figure~\ref{fig:Cls_kai} we compared the relativistic spectra obtained, as explained above, by assuming a standard flat model with cosmological constant $\Lambda$ (see the $\Lambda$CDM model in table~\ref{tab:cosmo}) to an open model with $\Omega_{\rm m}+\Omega_{\Lambda}=1-\Omega_K$ (see the o$\Lambda$CDM model in table~\ref{tab:cosmo}).
In figure \ref{fig:dyn_geom} and \ref{fig:dyn_geom_2} we will now compare the relativistic spectra from the o$\Lambda$CDM and $\Lambda$CDM models also to a spatially flat model including a non-adiabatic fluid, with $\Omega_{\rm m}+\Omega_{\Lambda}=1-\Omega_{\rm fld}$ (see the f$\Lambda$CDM model in table~\ref{tab:cosmo}).
The Hubble parameter reads
\be \label{eq:H_dyn_geom}
\frac{H}{H_0}=\sqrt{\Omega_{\rm m}(1+z)^3+\Omega_{\Lambda}+\Omega_{\rm X}(1+z)^2} \;,
\ee
where X refers either to curvature or to the fluid (in the latter case the metric in eq.~(\ref{metric}) has no curvature term).
We always set $\Omega_{\rm m}=0.3$.

Figure~\ref{fig:dyn_geom} shows the relative difference between the o$\Lambda$CDM and f$\Lambda$CDM models.
The dynamical evolution of the background is the same in both cases, hence differences in the spectra reflect the geometrical contributions induced by the curvature term.
Geometrical effects contribute substantially to the difference shown in figure~\ref{fig:Cls_kai}, and they  slightly increase at very large scales and for low redshift bins, especially in the auto-correlations.

In figure~\ref{fig:dyn_geom_2} we show the relative difference of the f$\Lambda$CDM and $\Lambda$CDM models.
In both cases the geometric contributions due to the curvature term are absent, hence the deviations provide information about the effect of the different evolution of the background.
Compared to  figure~\ref{fig:Cls_kai}, the amplitude of the deviations is less significant, but still non-negligible especially for the bin auto-correlations at larger redshift.
In the cross-correlations of the first bin  with other bins the dynamical contributions increase with multipole. Even though the cross correlation 1-2 also passes through zero at $\ell\simeq 13$, the difference does not change sign in this case, hence the relative difference simply shows a divergence since the zeros of both spectra do not exactly coincide. The corresponding plot of the difference between the curved and the fluid case shown in fig.~\ref{fig:dyn_geom}, however exhibits a change of sign (leading to the characteristic vertical line in the relative difference).
In the auto-correlations for the higher redshift bins the plateau in figure~\ref{fig:dyn_geom_2} is extended over larger multipoles and starts decreasing around the region of the acoustic peaks, to increase again later on (see e.g.~the bins 2-2 correlation).
In the case of the first bin auto-correlation, the increasing regime is reached already at the lower multipoles.
This is due to the fact that on large scales the $C_\ell$'s are roughly constant leading to a constant shift while on smaller scales power is increasing with $\ell$ such that a constant offset in the angular diameter distance leads to an increasing shift in the spectra. Since at low redshift also low $\ell$'s do not signify very large scales, this increase starts earlier for the lowest redshift bin.

Note that, in particular, the position of the acoustic peaks (see the amplitude of the fluctuations at $\ell \gtrsim 100$ in figures~\ref{fig:Cls_kai},~\ref{fig:dyn_geom} and~\ref{fig:dyn_geom_2}) is affected by a comparable amount by both dynamical and geometrical effects.
This is expected, since a different redshift-distance relation implies a shift in $\ell$ of the acoustic peaks: given a certain $z$, comoving distances for the open model o$\Lambda$CDM are larger than in a flat $\Lambda$CDM universe, and the acoustic scale of the peak subtends a smaller angle (larger multipole~$\ell$).

To conclude, both geometrical (i.e., changes in the metric) and dynamical effects (i.e., different cosmological parameters in eq.~(\ref{eq:H_dyn_geom})) are relevant and increase with the scale (except for the first bin auto-correlation in figure~\ref{fig:dyn_geom_2}).
In the configuration taken into account, geometrical effects dominate over the dynamical ones, which are nevertheless non-negligible especially if information about the precise acoustic peak locations is needed.

\subsection{Error forecasts}

In this section we forecast how future galaxy surveys will be able to measure the curvature parameter by using a new version of the code \class{}~\cite{Blas:2011rf,Lesgourgues:2013bra,DiDio:2013bqa}, that we modified to include the formalism described in the previous sections. The aim of this work is not to derive the best constraints for $\Omega_K$ achievable with future survey, but to understand how much information is encoded in the relativistic corrections and to determine the size of  the error in the curvature introduced by neglecting them.  We therefore consider a rather conservative set-up with only 5 redshift bins. As shown in~\cite{Asorey:2012rd,DiDio:2013sea}, a larger number of bins is required to extract all the 3-dimensional information from the survey.
With few bins, as we will show, the lensing convergence turns out to be the most relevant contribution neglected in a Newtonian galaxy clustering analysis. Including many thin redshift bins (which is in principle possible thanks to SKA's spectroscopic redshift determination $\sigma_z/(1+z)\sim0.001$) may enhance the effect of other local terms such as Doppler effects (see e.g.~\cite{Raccanelli:Doppler}), and a dedicated study of this may be worthwhile.

We assume specifications consistent with a SKA-like survey \cite{Santos:2015hra,Camera:2014bwa}, described in Appendix \ref{sec:forecast} together with our Fisher matrix formalism. There, we also specify our base values of cosmological parameters which agree roughly with the Planck results~\cite{Planck:2015xua}.
We expect our forecasts to hold qualitatively also for an Euclid-like photometric survey, the main differences being given by a reduced sky coverage $f_{\rm sky}^{\rm Euclid}\approx0.3$---entering trivially in the covariance, eq.~(\ref{eq:err-clgt}), so that errors increase by $\left[f_{\rm sky}^{\rm SKA}/f_{\rm sky}^{\rm Euclid}\right]^{1/2}$, without relative changes---and by larger redshift uncertainties, whose effect is partially reduced thanks to the large redshift bins; see, e.g., \cite{Montanari:2015rga} for a comparison of the lensing effect, main term of interest also in the present work, in the two cases SKA and Euclid.

In the rest of this Section we compare the information encoded in the Newtonian terms, in cosmic magnification and in the other relativistic effects. In particular, we want to study how the information beyond Newtonian terms may change the constraints on the curvature parameter $\Omega_K$. For this we shall analyse two questions:
1) to what extend constraints on the curvature parameter are degenerate with the accurate knowledge of relativistic effects?, and 2) how are the best-fits of cosmological parameters biased if relativistic effects are neglected? We will now address the first problem, and consider the second question in the next section.

\begin{figure}[t!]
\center
  \includegraphics[width=0.45\linewidth]{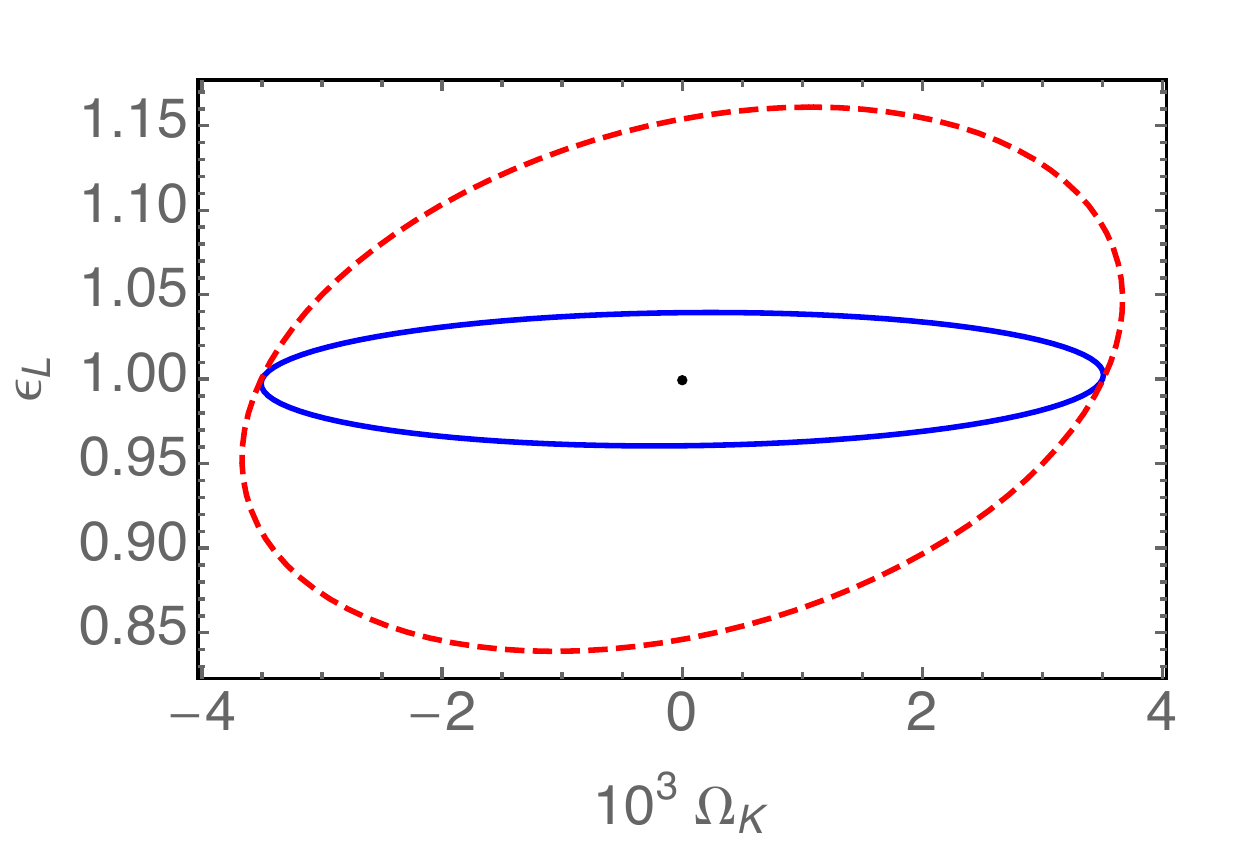}
  \includegraphics[width=0.45\linewidth]{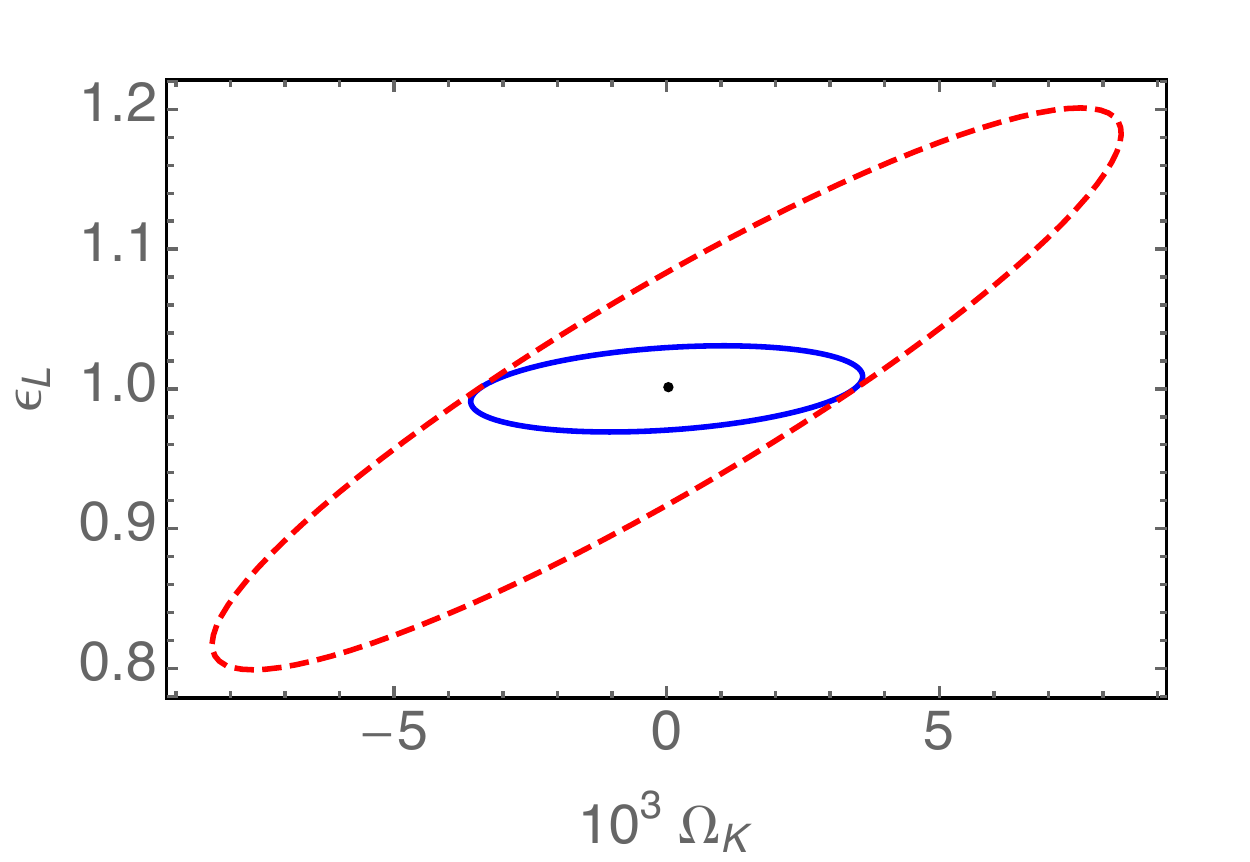}
    \caption{$1\sigma$ contour plot for $\Omega_K$ and $\epsilon_L$. In red (dashed) we consider only z-bin auto-correlation, while in blue (solid) we include also z-bin cross-correlation. In the left panel we adopt magnification bias for a SKA-like survey (i.e.~eq.~(\ref{sz_SKA})), while in the right panel we consider a constant magnification bias $s(z) = 1$. In both panels we have $f_\text{sky}=0.75 $ and we assume $\ell_\text{max} = 300$.  The parameters not shown in the plot are fixed to their fiducial values.
    }
  \label{fig:FOKepsilon}
\end{figure}
Cosmic magnification and other relativistic effects, if neglected, introduce a systematic error in the data analysis. To estimate this error, we perform a Fisher analysis (see Appendix~\ref{sec:forecast} for more details) and we compare the forecasted error on  $\Omega_K$ if we do include cosmic magnification with respect to a complete marginalization of the cosmic magnification amplitude.
As in~\cite{Montanari:2015rga,Alonso:2015uua}, we introduce a new parameter,~$\epsilon_L$, which describes the amplitude of cosmic magnification, by replacing\footnote{We remark that the parameter $\epsilon_L$ is not included in the public released version of \class{}. Nevertheless, the user can easily implement it following the details presented in Appendix~\ref{s:transfer}.} $\Delta_{\ell}^{\mathrm{Len}}$ with $ \epsilon_L \Delta_{\ell}^{\mathrm{Len}}$ in eq.~(\ref{eq:D_rel}).
We treat $\epsilon_L$ on the same footing as other cosmological parameters in the Fisher analysis, namely we include it both in the derivatives of the spectra and on the covariance matrices.
As we see in figure~\ref{fig:FOKepsilon}, the lensing information can be highly degenerate with $\Omega_K$ if we consider only z-bin auto-correlations (red contours). The amount of degeneracy (i.e., the difference between marginalized and fixed 1-dimensional constraints on $\Omega_K$) is very sensitive to the magnification bias parameter $s(z)$, which has to be known with  good accuracy in order not to systematically affect the constraints on the curvature parameter. The degeneracy seems to be completely broken once we include z-bin cross-correlations. Hence, a full tomographic analysis will not introduce a systematic change in the forecasted error on $\Omega_K$. This shows how crucial is to use the radial information to be able to disentangle the clustering signal from the lensing magnification. Clearly, the constraints on the lensing potential increase significantly when we include z-bin cross-correlation, this is discussed in detail in~\cite{Montanari:2015rga}. On the other hand, the constraints on the curvature parameter come mainly from the transversal modes. Indeed by adding the radial mode, and accounting properly for the cosmic magnification as forecasted for SKA, the error does not change.

\begin{figure}[t!]
\center
  \includegraphics[width=0.45\linewidth]{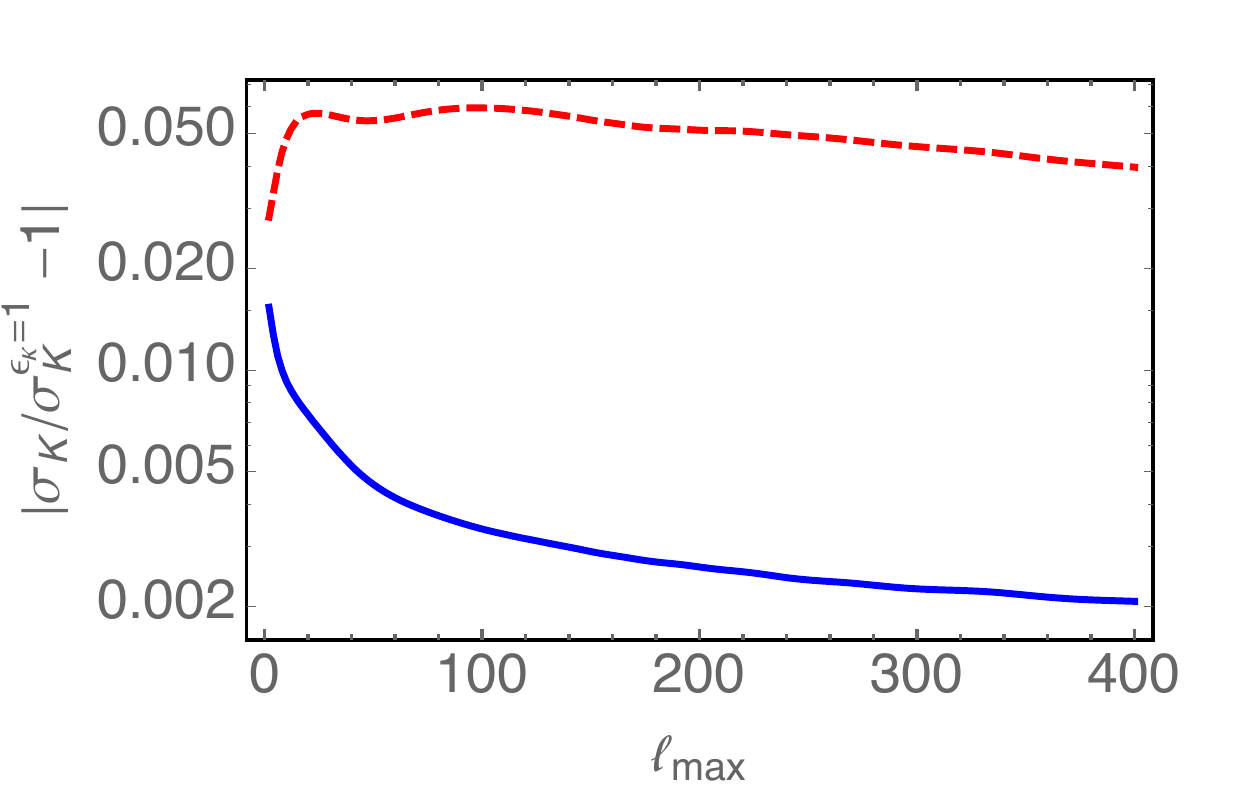}
  \includegraphics[width=0.45\linewidth]{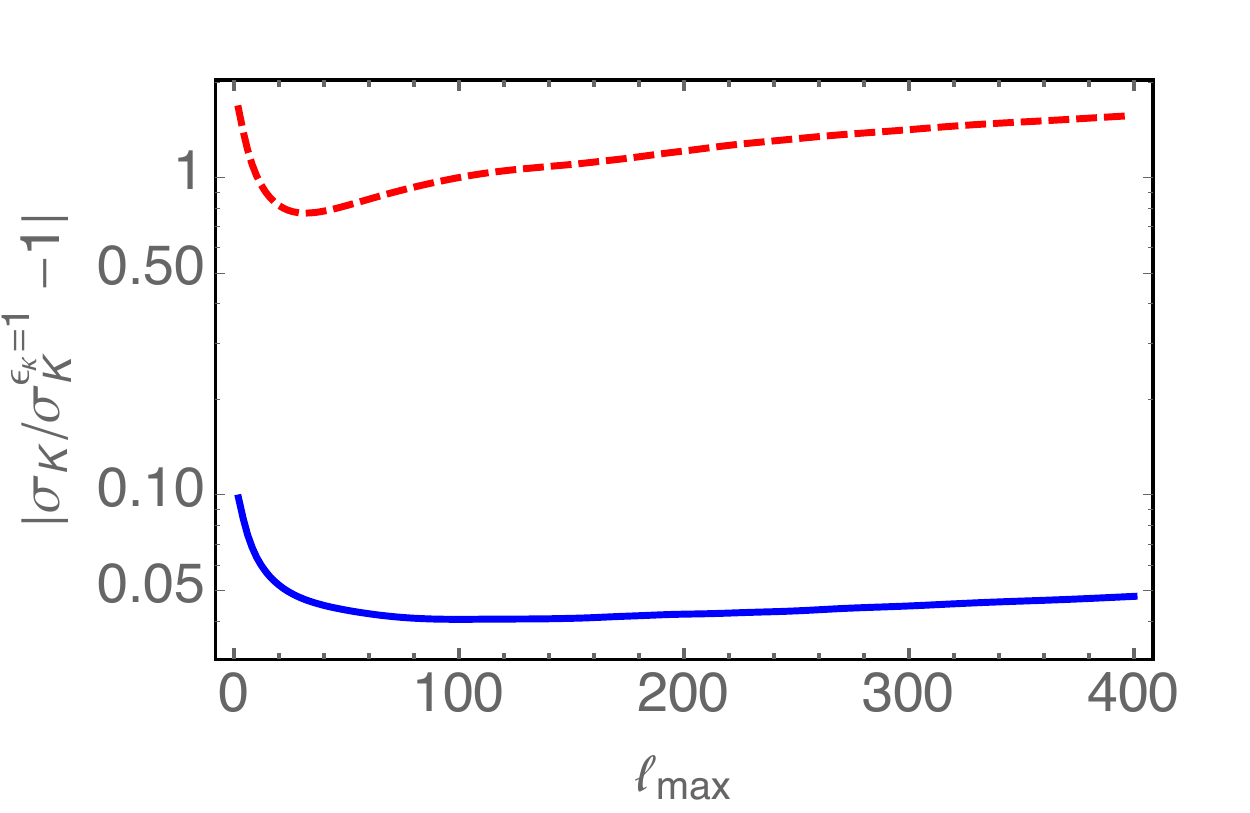}
    \caption{We show the relative difference between the forecasted error on $\Omega_K$ for a constant cosmic magnification amplitude and the case marginalized over $\epsilon_L$ as a function of the non-linear scale $\ell_\text{max}$. We consider the magnification bias for a SKA-like survey (left panel) and with a constant magnification bias $s(z) = 1$ (right panel). Red (dashed) lines include z-bin auto-correlation only, while blue (solid) lines consider as well z-bin cross-correlation.}
  \label{fig:FOKepsilon_ellmax}
\end{figure}
In figure~\ref{fig:FOKepsilon_ellmax} we show the relative difference between the forecasted error on $\Omega_K$ for a constant cosmic magnification amplitude and the case marginalized over $\epsilon_L$ as a function of the non-linear cut-off $\ell_\text{max}$. We notice that the degeneracy between cosmic magnification and  curvature is not broken by going to smaller scales, if we do not take into account radial correlations, i.e., z-bin cross-correlations.
We also show the case for $s(z) = 1$ (instead of the redshift dependent function compatible with SKA), which proves the importance of modeling correctly magnification bias.

\begin{figure}[t!]
  \center
  \includegraphics[width=0.45\linewidth]{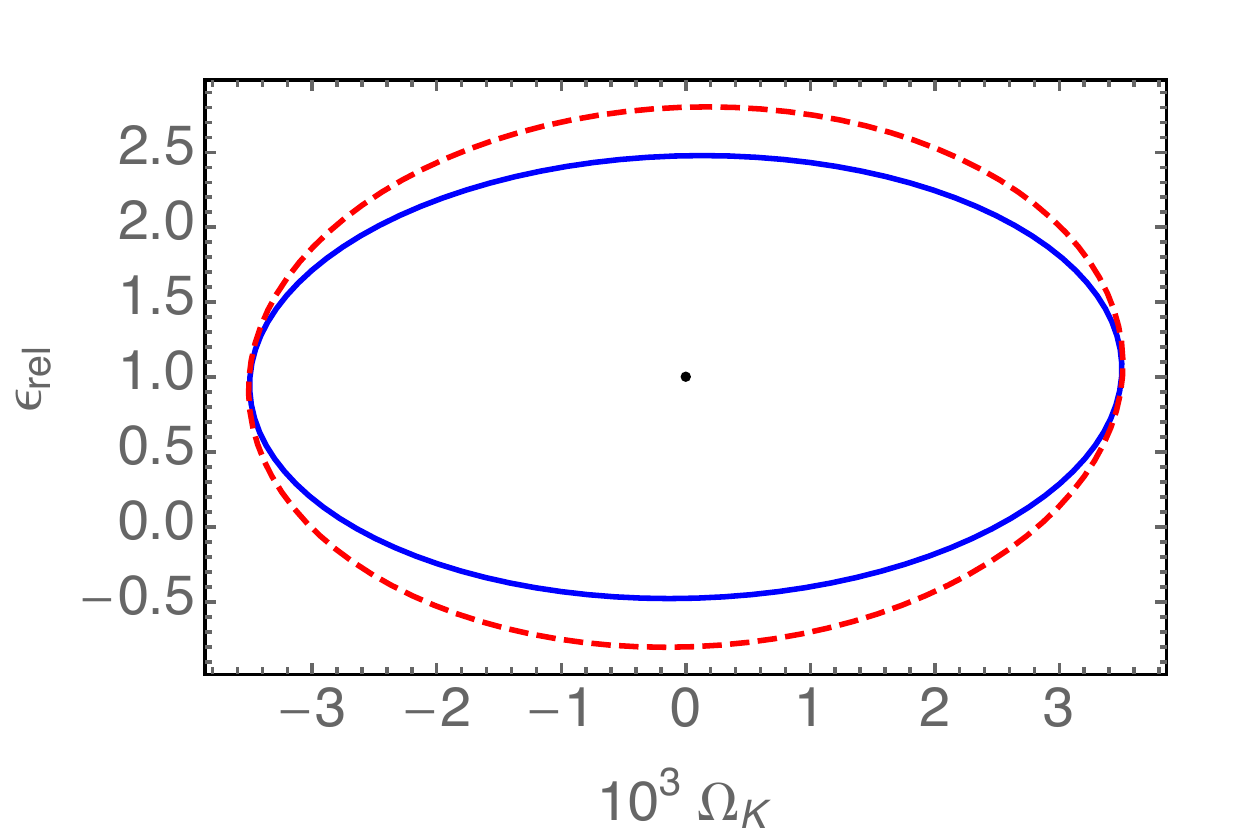}
    \caption{$1\sigma$ contour plot for $\Omega_K$ and $\epsilon_\text{rel}$. In red (dashed) we consider only z-bin auto-correlation, while in blue (solid) we include also z-bin cross-correlations. We adopt magnification bias for a SKA-like survey (i.e.~eq.~(\ref{sz_SKA})). In both panels we have $f_\text{sky}=0.75 $ and we assume $\ell_\text{max} = 300$.
    }
  \label{fig:FOKepsilonGR}
\end{figure}
%
We repeat the same approach to study the impact of the other relativistic effects (apart from the cosmic magnification studied so far) on constraints of the curvature parameter $\Omega_K$. We now replace in equation~(\ref{eq:D_rel}) $ \sum_{a=1}^2 \Delta_{\ell}^{\mathrm{D}a} + \sum_{a=1}^5 \Delta_{\ell}^{\mathrm{G}a}$ with $ \epsilon_\text{rel}\left(  \sum_{a=1}^2 \Delta_{\ell}^{\mathrm{D}a} + \sum_{a=1}^5 \Delta_{\ell}^{\mathrm{G}a} \right)$. In figure~\ref{fig:FOKepsilonGR} we see that these other relativistic effects and the curvature parameter are completely independent, either with or without adding z-bin cross-correlations. Contrary to the cosmic magnification, we notice that relativistic effects are poorly constrained with our conservative setting based on only 5 z-bins. Because of this very low constraining power on $\epsilon_\text{rel}$, the forecasted errors on $\Omega_K$ are insensitive to the amplitude of relativistic effects. This result agrees with previous analysis on different cosmological parameters, which have shown that $\epsilon_\text{rel}$ can be detected only through multi-tracer techniques if few redshift bins are considered~\cite{Yoo:2012,Alonso:2015sfa,Fonseca:2015laa}, whereas cosmic magnification is relevant also in a single tracer analysis~\cite{Montanari:2015rga}.

\subsection{Bias on the best-fit}

Now we study how neglecting relativistic integrated terms can bias the best fit measurement of the curvature parameter. A complete investigation of the shift in the best fit of the measurement would require an MCMC analysis~\cite{Cardona:2016qxn}. However, if the difference between the two models is small, we can Taylor expand and, to leading order in the systematic effect, we can use eq.~(\ref{eq:shift}).
We focus on measurements of curvature jointly with one other parameter, keeping the remaining parameters at their base values, our fiducial cosmology.
As second parameter we consider both the dark matter energy density parameter $\Omega_{\rm cdm}$, and the dark energy equation of state parameter $w_0$ (see appendix~\ref{sec:forecast} for more details about the fiducial cosmology and specifications).

\begin{figure}[t!]
  \center
  \includegraphics[width=0.45\linewidth]{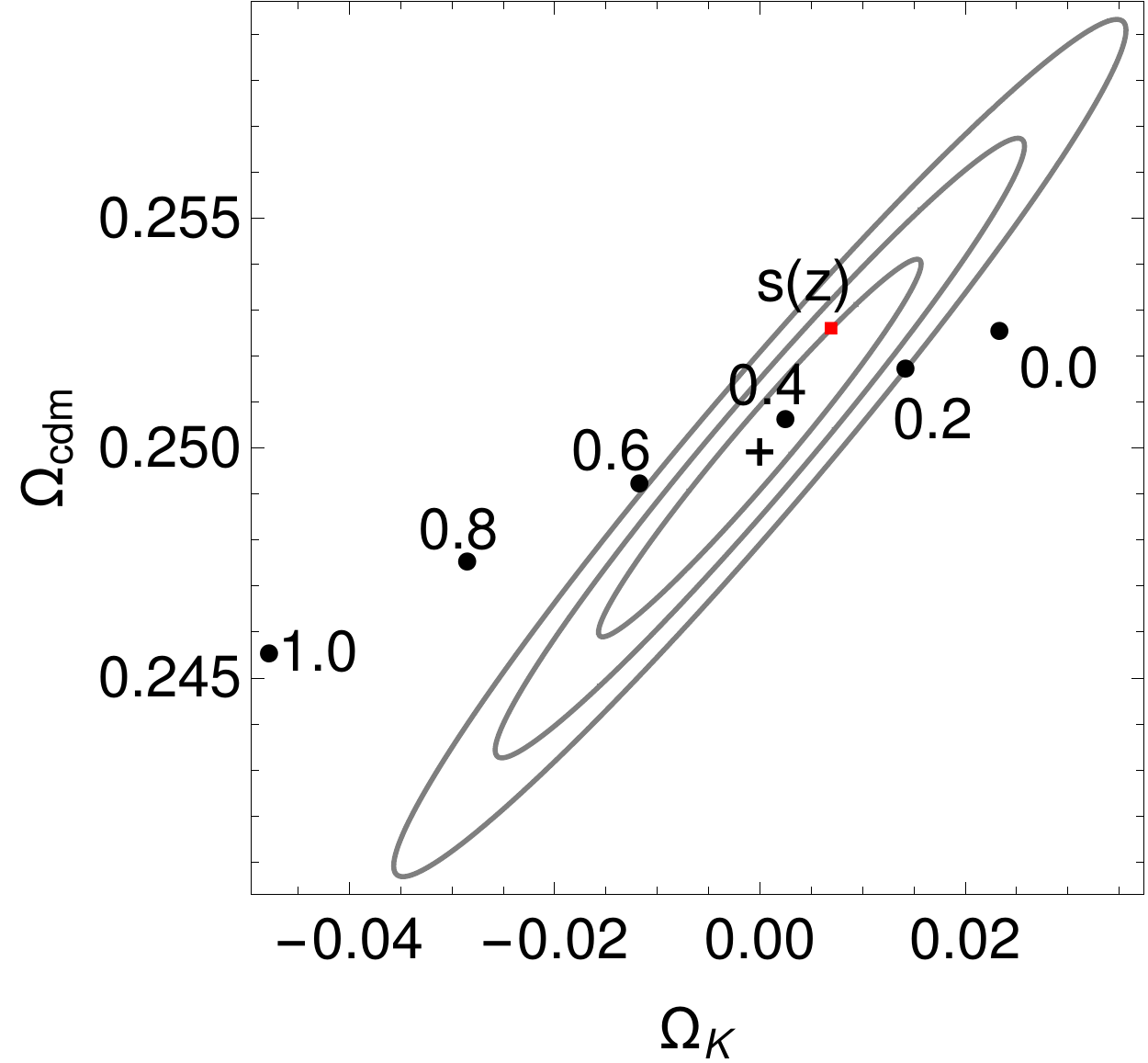}
  \quad
  \includegraphics[width=0.45\linewidth]{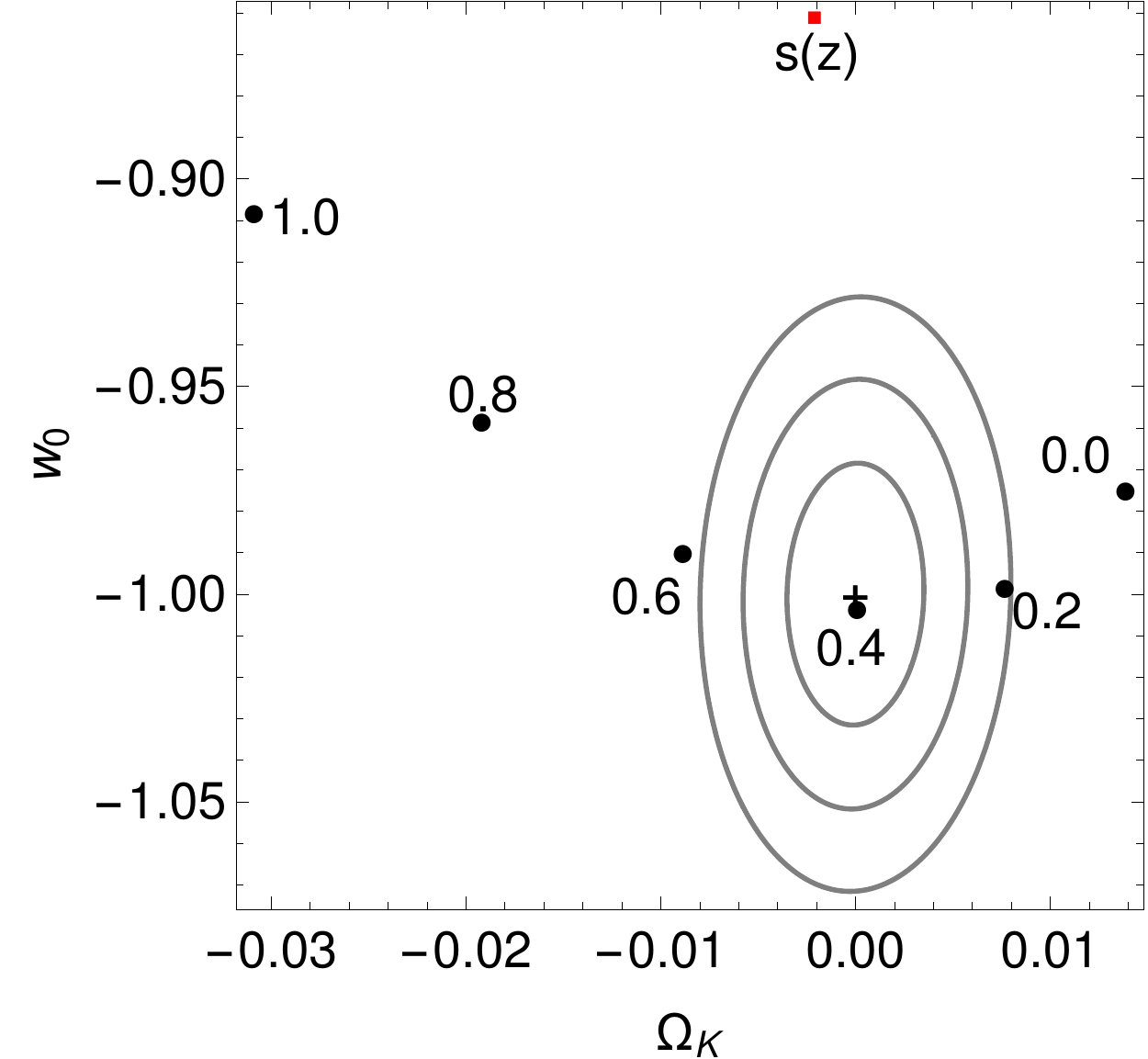}
  \caption{
    Black dots denote shifts in the best fit measurement due to neglecting relativistic terms, for the joint analysis of the curvature parameter and the CDM density parameter (left panel) or the DE equation of state parameter (right panel) due to neglecting relativistic effects. Different constant values of the magnification bias $s$ are indicated, whereas the label `$s(z)$' (red square) refers to the redshift-dependent case depicted in figure~\ref{fig:SKA_spec}. The fiducial parameters are indicated by a cross, which corresponds to the expected best-fit values if the analyses where done consistently including all relativistic terms.
    We plot $1\sigma$, $2\sigma$ and $3\sigma$ contours obtained by a Newtonian analysis, for $\ell_{\rm max}=300$.
  }
  \label{fig:shift_s}
\end{figure}

\begin{figure}[t!]
  \center
    \includegraphics[width=0.42\linewidth]{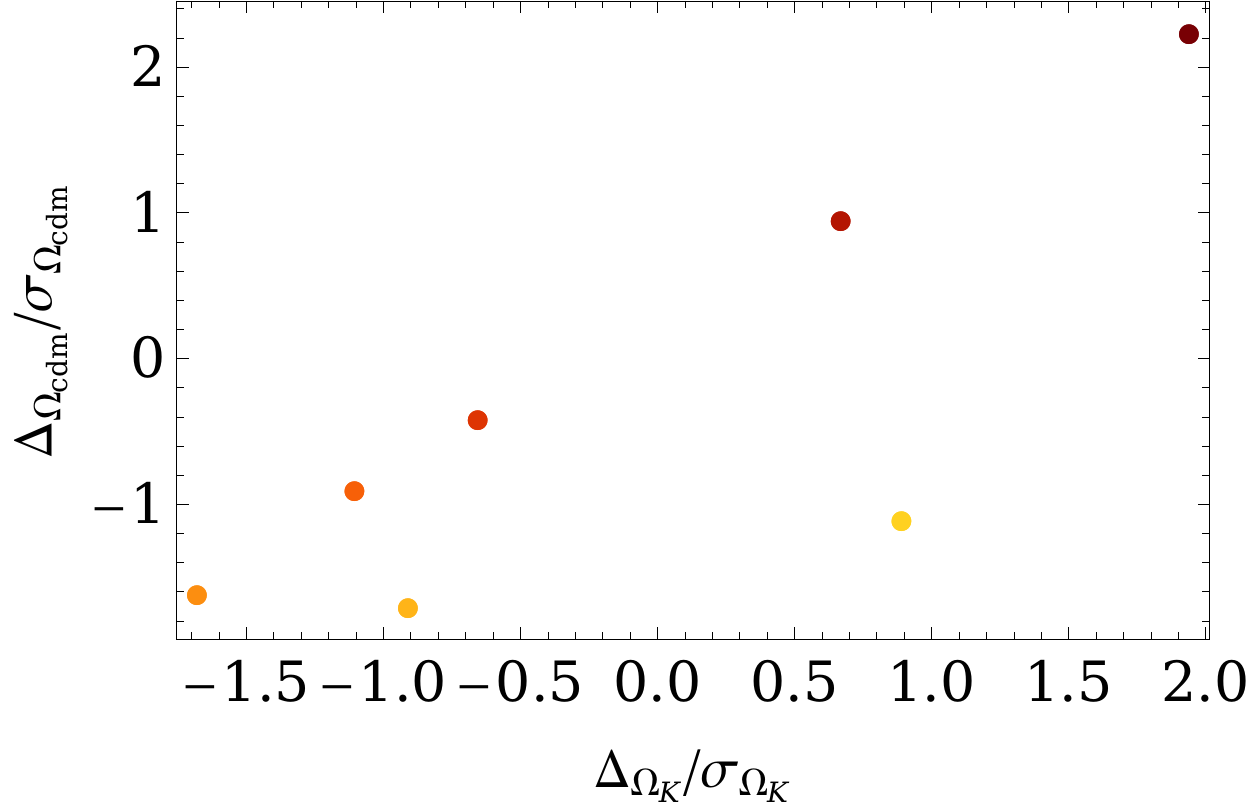}
    \quad
    \includegraphics[width=0.4\linewidth]{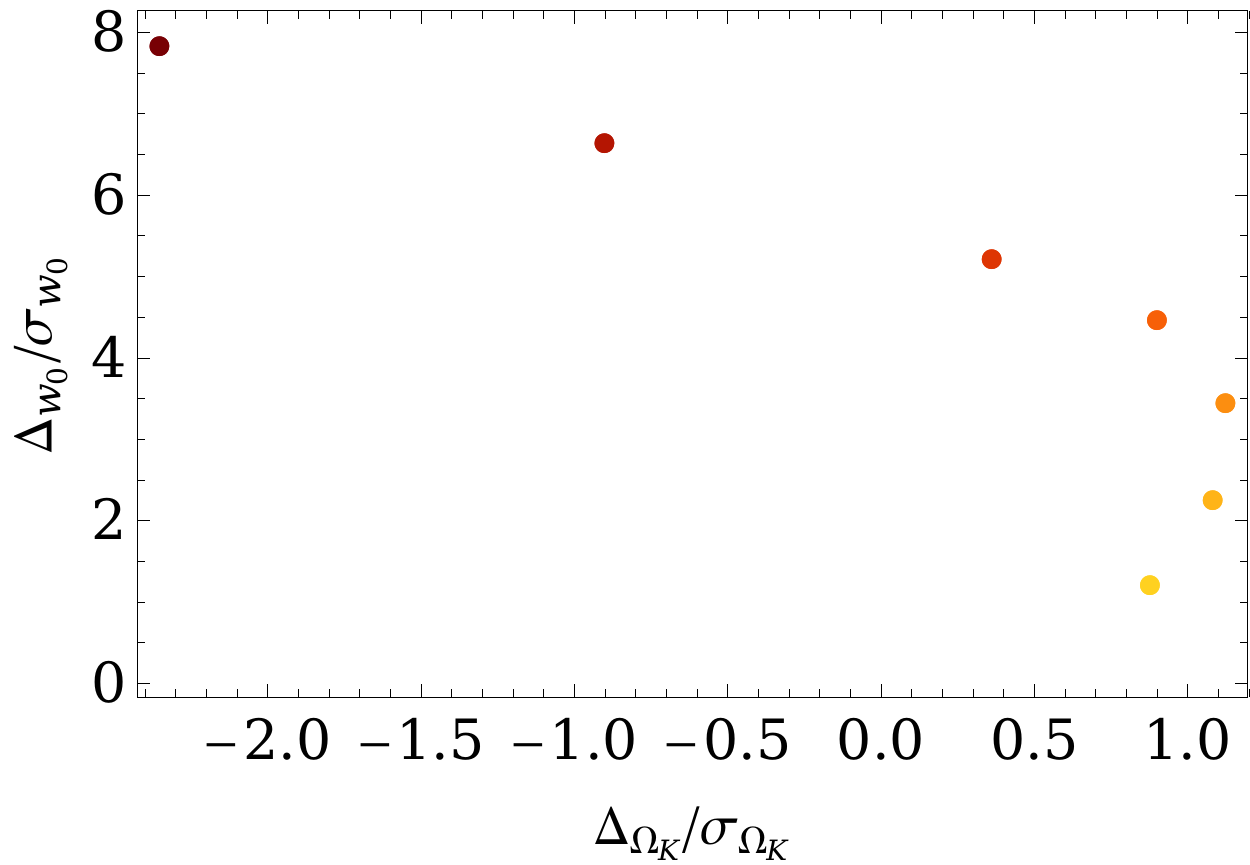}
    \quad
    \includegraphics[width=0.08\linewidth]{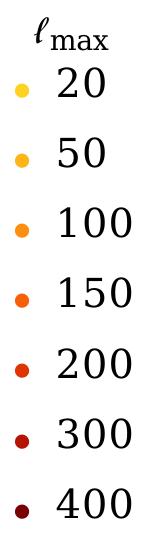}
    \caption{
      Shifts in the best fit parameters relative to the standard errors for the joint analysis of the curvature parameter and CDM parameter (upper panel) or DE equation of state (bottom panel) due to relativistic effects. Magnification bias $s(z)$ has been fixed to the function shown in figure~\ref{fig:SKA_spec}.
      Different dots correspond to different values (lower for lighter dots, larger for darker dots) of the maximum multipole $\ell_{\rm max}$ considered in the analysis.
    }
  \label{fig:shift_l}
\end{figure}

Figure~\ref{fig:shift_s} shows the bias in the estimation of best-fit cosmological parameters obtained with the Fisher formalism outlined in eq.~(\ref{eq:shift}).
Here, in order to estimate the systematic error due to neglecting relativistic terms, we include in the analysis all redshift bin cross-correlation both in the ``Newtonian'' and in the ``full relativistic'' analysis.
Contours correspond to the $1\sigma$, $2\sigma$ and $3\sigma$ intervals obtained from the Fisher matrix in the {Newtonian} approximation.
Red squares correspond to the shift expected by assuming the full redshift-dependent magnification bias $s(z)$ showed in figure~\ref{fig:SKA_spec}.
We consider the joint constraints on $\{\Omega_K,\Omega_{\rm cdm}\}$ and on $\{\Omega_K,w_0\}$, where in each case the other standard $\Lambda$CDM parameters are kept fixed to their fiducial values (i.e., in each plot we consider a $2\times2$ Fisher matrix).
Interestingly, the joint constraint of $\{\Omega_K,w_0\}$ shows a much larger systematic shift than the one of the pair $\{\Omega_K,\Omega_{\rm cdm}\}$ (this is investigated more in detail below, see figure~\ref{fig:shift_l}).

We also estimate the shifts by assuming several constant values of magnification bias $s$, indicated in figure~\ref{fig:shift_s} next to the black dots.
The case $s=0.4$ always shows a small shift. In this case, the lensing magnification term vanishes and the shift comes entirely from the other relativistic terms.
This confirms the fact that, in our tomographic configuration including few relatively large redshift bins, the lensing term is the main non-Newtonian contribution.
Indeed, as it can be seen from eq.~(\ref{eq:NC_obs}), the pre-factor  $(2-5s)$ of the lensing integral vanishes for $s=0.4$.
We verified that the main remaining contribution to the shift for $s=0.4$ is due to Doppler terms.
The direction of the shift along the $\Omega_K$ direction is due to the fact that a negative $(2-5s)$ factor corresponds to observing less volume at a given angular separation, hence it corresponds to a positive curvature $K>0$ (i.e., $\Omega_K<0$). This increase in the curvature, however leads to a smaller angular diameter distance which is compensated by a lower $\Om_\text{cdm}$ and correspondingly higher $\Om_\La$. In the right panel, where $\Om_\text{cdm}$ is fixed and $\Om_\La$ increases with curvature, this leads to an increase in the angular diameter distance which can be compensated partially with and increase in $w_0$.
Of course this is not the full explanation as number counts are not only sensitive to the  angular diameter distance. Especially, the $s(z)$ specification for SKA (red square), which is small at low redshift but tends to $1$ at large redshift prefers only very little curvature but a $w_0$ significantly larger than $-1$.

We conclude that if the Newtonian analysis is assumed to be the correct model instead of the relativistic one (including at least also lensing), important systematic shifts in the best-fit values are to be expected depending on the survey specifications.
We cannot be more quantitative at this point, since the Fisher analysis presented here actually looses its validity when the shifts become significantly larger than $1\si$ which happens in our analysis.

In figure~\ref{fig:shift_l} we plot the systematic shift in the best-fit of the given parameter, computed with eq.~(\ref{eq:shift}), relative to the marginalized standard error estimated as $\sigma_{\alpha}=\sqrt{\left[(F^{\rm nwt})^{-1}\right]_{\alpha\alpha}}$, where $\left[(F^{\rm nwt})^{-1}\right]_{\alpha\alpha}$ is a diagonal element of the inverted Newtonian Fisher matrix. As in figure~\ref{fig:shift_s}, for each plot we fix the parameters that are not being considered to their fiducial values, hence in each case we only have a $2\times2$ Fisher matrix.
We assume a redshift-dependent magnification bias consistent with figure~\ref{fig:SKA_spec}.
The different dots show the shifts obtained for different multipole cut-off values $\ell_{\rm max}$ used in the analysis (we recall that in  figure~\ref{fig:shift_l} we consider $\ell_{\rm max}=300$).
Lighter dots correspond to lower $\ell_{\rm max}$, while darker dots include more non-linear scales.
A more rigorous analysis would be necessary to treat consistently non-linear scales (see e.g.~\cite{DiDio:2013sea,Montanari:2015rga}), that we are here including for the lowest redshift bins and highest multipoles.
However, here we want to analyze qualitatively the dependence of the systematic shift on the largest scales included in the analysis. For a quantitative estimation an MCMC analysis is advisable \cite{Cardona:2016qxn} especially to confirm large shifts $\Delta_\theta\gg\sigma_\theta$, for which our formalism is not self-consistent.
In fact,  the Fisher matrix analysis can be trusted only for a small systematic bias, within the $1\si$ errors. When the shift is large, $\De_\theta/\si_\theta>1$ we can only conclude that there is a significant shift but we cannot trust its amplitude which is obtained by a first order Taylor expansion.
When only large scales are included in the analysis, the shifts are relatively modest, and they become significant as smaller scales are included.
This is in part due to the fact that, including more scales, the errors  shrink and a given shift becomes more relevant.
But also, the shift itself increases considerably, since the lensing integral becomes more relevant at larger multipoles \cite{DiDio:2013sea,Montanari:2015rga}.
Instead, the shift increases significantly (and monotonically) in the $w_0$ direction, which is then particularly sensitive to the lensing effect. This shows also that
radial correlations are especially useful to constrain  the evolution of dark energy.

\section{Conclusions}
\label{sec:conclusions}
In this work we generalize the observed galaxy number counts to spatially curved  geometries. We include all relativistic effects to first order in perturbation theory. We implemented this new formalism in the new release {\sc v2.5.0} of the publicly available code \class{}~\cite{Blas:2011rf,Lesgourgues:2013bra,DiDio:2013bqa}, to compute efficiently and accurately the redshift dependent angular power spectra.

We have compared the constraining power on $\Omega_K$ coming from its geometrical nature and from its dynamical effects and have shown that the geometrical part is prevalent.
Then, we used the fully relativistic formalism to investigate how constraints on the spatial curvature of the universe coming from future galaxy surveys differ when using the proper relativistic formalism rather than the standard Newtonian approach.
We showed how the error on $\Omega_K$ depends on relativistic effects and how the inferred value changes when we properly account for all the relativistic corrections. We verified that the most relevant relativistic contribution in our configuration (compatible with a SKA-like survey divided in 5 bins) is  cosmic magnification.
We have found that including radial information, i.e.~z-bin cross-correlations data, is in general crucial and significantly reduces the errors and the degeneracy between spatial curvature and cosmic magnification. However, the significance of this degeneracy strongly depends on the magnification bias $s(z)$.
Actually, for the function predicted for SKA it is relatively mild.

We then have studied the bias in the inferred best-fit value of $\Omega_K$, obtained when neglecting magnification or the other relativistic terms. While the bias from the other relativistic terms is much smaller than the errors, the shift from magnification can become substantial, depending on the value of the magnification bias $s$.
When deriving constraints on $\Omega_K$ from future LSS surveys, it is therefore very important to include magnification and to have good knowledge of the magnification bias $s(z)$ of the survey.

This work can be regarded as a contribution to the  general effort in setting up a precise theoretical modeling for analyzing galaxy clustering using future surveys and it provides the community with an improved tool to better understand galaxy clustering at the largest scales.
With the use of a formalism including large-scale effects, wide and deep future galaxy surveys can avoid any theoretical bias introduced by neglecting relativistic effects.

Furthermore, current measurements of the curvature parameter are not yet at cosmic variance limit. The effort of reaching the level of accuracy needed to test some very fundamental theories about the nature of the Universe, and to provide observational constraints on models inspired by the string landscape, can benefit of a full relativistic description.

\acknowledgments{
ED is supported by the ERC Starting Grant cosmoIGM and by INFN/PD51 INDARK grant.
RD acknowledges financial support by the Swiss National Science Foundation.
FM acknowledges the hospitality of the Department of Physics \& Astronomy, Johns Hopkins University, where part of this work was developed.
This work was supported at JHU by NSF Grant No. 0244990, NASA NNX15AB18G, the John Templeton Foundation, and the Simons Foundation.
}

\appendix
\section{Scalar harmonic functions}
\label{s:harmonic}

Here we give the explicit form of the orthogonal functions defining a basis for expansion of scalars $X(\bn,z)$ in wavenumber space.
The expansion is performed in spherical coordinates and allows for an arbitrary curvature $K$.
The derivation follows the works of \cite{Abbott:1986ct,Hu:1997mn,DurrerBook,Tram:2013xpa,Lesgourgues:2013bra}.
For a possible generalization to vector and tensor modes see, e.g., \cite{Dai:2012bc}.

In this appendix we use the same notation as in the \class{} code~\cite{Blas:2011rf,Lesgourgues:2013bra,DiDio:2013bqa}, where the spatial part of the metric in eq.~(\ref{metric}) is written as
\be
\label{metric_class}
\gamma_{ij} dx^i dx^j = |K|^{-1} \left[ d\chi^2 + \sin_K^2\chi\left( d\theta^2 + \sin^2\theta d\varphi^2  \right)  \right] \;,
\ee
and the rescaled radial coordinate is $\chi=\sqrt{|K|}(\tau_0-\tau)=\sqrt{|K|}r$.
We defined
\bea
\label{sinK}
\sin_K \chi = \left\{ \begin{array}{cc}
\sin \left( \chi \right) & \text{for} \ K>0 \\
\sinh \left( \chi \right) & \text{for} \ K<0
\end{array}
\right. \, ,
\eea
which is related to our previous notation by $\CHI(r)=\sin_K(\chi)/\sqrt{|K|}$.
The flat-limit is consistently obtained taking $|K|^{-1/2} \sin_K\chi \xrightarrow[]{K\to0} r$.
We also define
\bea
\cot_K \chi = \left\{ \begin{array}{cc}
\cot \left( \chi \right) & \text{for} \ K>0 \\
\coth \left( \chi \right) & \text{for} \ K<0
\end{array}
\right. \,,
\eea
with flat-limit $|K|^{1/2}\cot_K\chi \xrightarrow[]{K\to0} 1/r$.

We look for eigenfunctions $Q_{\bk}(\bx)$ of the covariant Laplacian $\Delta_K$ given the metric (\ref{metric_class}).
These are the solutions of the Helmotz equation
\be
\label{helmotz}
(\Delta_K + k^2) Q_{\bk}(\bx) = 0\;.
\ee
Introducing the generalized wavenumbers
\be
q=\sqrt{k^2+K}\;, \qquad \ka=q/\sqrt{|K|} \;,
\ee
the spectrum is complete for
\bea
\begin{array}{rlc}
\ka & =3,4,5,\ldots & \text{for } K>0 \\
    & \geq 0 & \text{ for } K\leq0 \;.
\end{array}
\eea
For possible issues of super-curvature modes in an open Universe, see \cite{Liddle:2013czu}.
In the following we change notation between $k$. $q$ and $\ka$ as most convenient.

In flat space $K=0$ the equation is solved\footnote{
    We use unitary Fourier transform conventions. For $K=0$ we define $Q_{\bf k} = \frac{1}{(2\pi)^{3/2}}e^{i {\bf k}\cdot{\bf x}}$, the power spectrum is $\langle \Delta({\bf k}) \Delta({\bf k}')  \rangle = \delta({\bf k}+{\bf k}') P(k)$ and the variance per logarithmic wavenumber $\mathcal{P}(k) = \frac{k^3}{2\pi^2}P(k)$, such that in real space $\xi(r) \equiv \langle \Delta({\bf x}) \Delta({\bf x} + r\bn) \rangle = \int_0^\infty\frac{dk}{k}j_0(rk){\cal P}(k)$.
}
 by $Q_{\bk}(\bx)=\frac{1}{(2\pi)^{3/2}}\exp(i \bk\cdot\bx)$, which satisfy the orthonormality relation
\be
\label{eq:closure_flat}
\int d^3 \bx\; Q_{\bk}(\bx) \left[ Q_{\bk'}(\bx)  \right]^* =  \delta(\bk-\bk') \;,
\ee
where the star denotes complex conjugation, and $\delta(\bk-\bk')$ is the Dirac delta.
Eq.~(\ref{helmotz}) is also solved by the functions $Q_{k\ell m}(\bx)=j_{\ell}(kr)Y_{\ell m}(\hat\bx)$, where  $\hat\bk \equiv \bk/k$, $Y_{\ell m}(\bn)$ are spherical harmonics and $j_{\ell}(kr)$ are spherical Bessel functions.
Since spherical harmonics form a complete system of functions on the sphere, there must exist an expansion $\exp(i \bk\cdot\bx) = \sum_{\ell,m} c_{\ell m} Q_{k\ell m}(\bx)$.
The computation of the coefficient $c_{\ell m}$, see e.g.~\cite{DurrerBook}, leads to:
\be
\label{exp_harmonic}
Q_{\bk}(\bx) = \frac{1}{(2\pi)^{3/2}} \exp(i \bk\cdot\bx)
= \sqrt{\frac{2}{\pi}} \sum_{\ell,m} i^l j_{\ell}(kr) Y_{\ell m}(\bn) Y^*_{\ell m}(\hat\bk) \;,
\ee
where we denote $\bk\cdot\bx = kr \hat\bk\cdot\hat\bx$.
We further remark that spherical Bessel functions satisfy the orthogonality relation
\be
\label{eq:orth_jl}
\int_0^{\infty} dr\, r^2j_{\ell}(kr)j_{\ell}(k'r) = \frac{\pi}{2} \frac{1}{k^2} \delta(k-k') \;.
\ee

The spatial part of each mode of oscillation in spherical coordinates is given by $j_{\ell}(kr) Y_{\ell m}(\bn)$, which as said satisfies eq.~(\ref{helmotz}).
In our coordinates only the radial dependence of the metric, eq.~(\ref{sinK}), changes for $K\neq0$.
Hence we look for eigenfunctions of the Helmholtz equation of the form $Q_{\ka\ell m}=\Phi_\ell^{\ka}(\chi) Y_{\ell m}(\bn)$ (which then form a basis for expansion in spherical coordinates for each mode k).
The solution is given by hyperspherical Bessel functions:
\bea
\Phi_{\ell}^{\ka} (\chi) = \left\{ \begin{array}{cc}
\sqrt{\frac{\pi M_{\ka}^{\ell} }{2 \ka^2 \sin\chi}} P_{-1/2+\ka}^{-1/2-\ell}(\cos\chi)  & \text{ for } K>0 \\
 & \\
\sqrt{\frac{\pi N_{\ka}^{\ell} }{2 \ka^2 \sinh\chi}} P_{-1/2+i\ka}^{-1/2-\ell}(\cosh\chi)  & \text{ for }  K<0
\end{array}
\right. \, ,
\eea
where
\be
M_{\ka}^{\ell} = \prod_{n=0}^{\ell} (\ka^2-n^2)\;, \qquad N_{\ka}^{\ell} = \prod_{n=0}^{\ell} (\ka^2+n^2) \,,
\ee
and $P_L^M(\mu)$ are associated Legendre functions~\cite{abramowitz}.
The normalization is chosen such that for $K\to0$ we recover $\Phi_{\ell}^{\ka}\to j_{\ell}(\ka\chi)=j_{\ell}(kr)$.
This leads to the orthogonality relation
\be
\label{eq:orth_Qnlm}
\int d^3\bx\; Q_{\ka\ell m}(\bx) \left[ Q_{\ka'\ell' m'}(\bx)\right]^* = \frac{\pi}{2} \frac{1}{q^2} \delta(q,q') \delta_{\ell\ell'} \delta_{mm'} \,,
\ee
where the product $\delta_{\ell\ell'} \delta_{mm'}$ comes from the orthogonality of spherical harmonics, while the factor $\frac{\pi}{2} \frac{1}{q^2} \delta(q,q')$ is the generalization of eq.~(\ref{eq:orth_jl}) (where $j_{\ell}(kr)$ are replaced by $\Phi_{\ell}^{\ka}(\chi)$), for which we defined
\bea
\delta(q,q') = \left\{ \begin{array}{cc}
\delta_{qq'}  & \text{ Kronecker delta for } K>0 \\
\delta(q-q') & \text{ Dirac delta for } K\leq0
\end{array}
\right. \, .
\eea

Then, introducing a vector $\bq$ with length $q$ and direction $\hat\bq = \hat\bk$, the functions
\be
\label{eigenf_helmotz}
Q_{\bq}(\bx) \equiv \sqrt{\frac{2}{\pi}} \sum_{\ell,m} i^\ell \Phi_{\ell}^{\ka}(\chi) Y_{\ell m}(\bn) Y^*_{\ell m}(\hat{\bq}) \;,
\ee
are solutions of the Helmholtz equation (\ref{helmotz}), reduce to eq.~(\ref{exp_harmonic}) for $K=0$, and satisfy the closure relation
\be
\label{eq:closure_K}
\int d^3\bx\; Q_{\bq}(\bx) \left[Q_{\bq'}(\bx)\right]^* = \delta({\bf q},{\bf q}') \;,
\ee
where we used $\delta({\bf q},{\bf q}') = \frac{1}{q^2\sin\theta} \delta(q,q') \delta(\theta-\theta') \delta(\phi-\phi')$.

Finally, it is useful to introduce the following functions:
\bea
& \mathfrak{j}_{\ell}^{\ka}(\chi) \equiv \Phi_{\ell}^{\ka}(\chi) & \xrightarrow[K\to0]{ }  j_{\ell}(k(\tau_0-\tau)) \;, \label{eq:gen_jl} \\
& \mathfrak{j}_{\ell}^{\ka\prime}(\chi) \equiv \frac{\sqrt{|K|}}{k}\Phi_{\ell}^{\ka\prime}(\chi) & \xrightarrow[K\to0]{ }  j'_{\ell}(k(\tau_0-\tau)) \;, \label{eq:gen_jpl} \\
& \mathfrak{j}_{\ell}^{\ka\prime\prime}(\chi) \equiv \frac{|K|}{k^2}\Phi_{\ell}^{\ka\prime\prime}(\chi) & \xrightarrow[K\to0]{ }  j''_{\ell}(k(\tau_0-\tau)) \;, \label{eq:gen_jppl}
\eea
with simple flat space limits.
Following \cite{Lesgourgues:2013bra}, we also define
\bea
& \mathfrak{sin}_{K}(\chi) \equiv \frac{k}{\sqrt{|K|}}\sin_{K}(\chi) & \xrightarrow[K\to0]{ }  k(\tau_0-\tau) \;, \label{eq:sinKgen} \\
& \mathfrak{cot}_{K}(\chi) \equiv \frac{\sqrt{|K|}}{k} \cot_{K}(\chi) & \xrightarrow[K\to0]{ }  \frac{1}{k(\tau_0-\tau)} \;, \label{eq:cotKgen}
\eea

\section{The transfer functions for number counts}
\label{s:transfer}

The results of Appendix~{\ref{s:harmonic}} allow us to expand a scalar function $X(\bn,z)$ in terms of
\be
a_{\ell m}(z) = \int d\Omega_{\bn} Y^*_{\ell m}(\bn) X(\bn,z) \;,
\ee
where
\be
X(\bn,z) = \sumint d^3\bq\, X(\bq,z) Q_{\bq}(\bx)
= \sqrt{\frac{2}{\pi}} \sumint d^3\bq\, X(\bq,z)  \sum_{\ell,m} i^\ell \Phi_{\ell}^{\ka}(\chi) Y_{\ell m}(\bn) Y^*_{\ell m}(\hat\bq) \;,
\ee
where the symbol $\sumint$ indicates that for positive curvature the integral over $q$ has to be replaced by a sum.
At initial time the power spectrum of the curvature perturbation ${\mathcal R}$ is defined as $\left< {\mathcal R}(\bk) {\mathcal R}^*(\bk') \right> = P_{\mathcal R}(k) \delta(\bk-\bk')$, and the dimensionless initial power spectrum as
$\mathcal{P}_{\mathcal R}(k)=\frac{k^3}{2\pi^2}P_{\mathcal R}(k)$.
The power spectrum cross-correlating the scalars $X$ and $Y$ is then obtained in terms of transfer functions $\Delta_{\ell}^{X,Y}(q,z)$  as
\bea
\left< a^X_{\ell m}(z) \left(a^Y_{\ell m}(z')\right)^* \right> &=& 4\pi \sumint \frac{dq}{q} \Delta_{\ell}^X(q,z) \Delta_{\ell}^Y(q,z') \tilde{\mathcal{P}}_{\mathcal{R}}(q) \nonumber \\
&=& 4\pi \sumint \frac{dk}{k} \Delta_{\ell}^X(q,z) \Delta_{\ell}^Y(q,z') \mathcal{P}_{\mathcal{R}}(k) \;,
\eea
where we define $\tilde{\mathcal{P}}_{\mathcal{R}}(q) \equiv \frac{q^2}{q^2-K} \mathcal{P}_{\mathcal{R}}(k)$ and use $q^2=k^2+K$, hence $k dk= q dq$.
In the last integral the variable $q$ is a function of $k$.

All the terms in eq.~(\ref{eq:NC_obs}) can be expanded following the recipe outlined above.
The only terms requiring further discussion are those involving the velocity.
To treat them we recall that at linear order in perturbation theory \cite{DurrerBook} the decomposition of the $k$-mode of a vector field is of the form
\be
\label{eq:vec_Q}
v_i = V(k) \left( -\frac{1}{k} \partial_i Q_{\bq}(\bx) \right) \;,
\ee
were we only consider scalar modes.
Then, using $-\bn\cdot\nabla = \partial_r$ and $\chi=\sqrt{|K|}r$, the expansion of terms proportional to $\bn \cdot \bv$ reads for each $k$-mode can be written as
\be
\bn\cdot\bv = \frac{\Theta(k)}{k} \left( \frac{\sqrt{|K|}}{k} \frac{\partial}{\partial \chi}Q_{\bq}(\bx) \right) \;,
\ee
where we define $\Theta(k)\equiv k V(k)$ as in \cite{Ma:1995ey} to match the notation of the \class{} code.
In the same way we can write
\be
\partial_r(\bn\cdot\bv) = \Theta(k) \left( \frac{|K|}{k^2} \frac{\partial^2}{\partial \chi^2}Q_{\bq}(\bx) \right) \;.
\ee
Expressing the eigenfunctions $Q_{\bq}(\bx)$ in terms of hyperspherical Bessel functions $\Phi_{\ell}^{\ka}\equiv\mathfrak{j}_{\ell}^{\ka}$ as in eq.~(\ref{eigenf_helmotz}), the terms $\bn\cdot\bv$ and $\partial_r(\bn\cdot\bv)$ involve the functions $\mathfrak{j}_{\ell}^{\ka\prime}$ and $\mathfrak{j}_{\ell}^{\ka\prime\prime}$, respectively, as one sees by comparing the equations written above with eq.~(\ref{eq:gen_jl}-\ref{eq:gen_jppl}).
Finally, the velocity term proportional to $v$, where $\bv = -\nabla v$, is a scalar and is expanded in terms of $\mathfrak{j}_{\ell}^{\ka}$.
 Comparing with eq.~(\ref{eq:vec_Q}) we obtain for each $k$-mode  $v=v(k)Q_{\bq}(\bx)$
\be
v(k)=V(k)/k \;.
\ee
Adding all the terms of eq.~(\ref{eq:NC_obs}) together we finally
express the transfer function of $\Delta({\bf n},z)$  as the sum of the following terms:

\begin{eqnarray}
\Delta_{\ell}^{\mathrm{Den}_i} &=& \int_0^{\tau_0} d\tau W_i \, b(z) S_\mathrm{D} \, \mathfrak{j}_{\ell}^{\ka} \nonumber \\
\Delta_{\ell}^{\mathrm{Red}_i} &=& \int_0^{\tau_0} d\tau \, W_i \left( \frac{1}{aH} \right) S_{\Theta} \, \mathfrak{j}_{\ell}^{\ka\prime\prime} \nonumber \\
\Delta_{\ell}^{\mathrm{Len}_i} &=& \ell(\ell+1) \int_0^{\tau_0} d\tau \, W^\mathrm{L}_i \,  S_{\Phi+\Psi} \, \mathfrak{j}_{\ell}^{\ka} \nonumber \\
\Delta_{\ell}^{\mathrm{D}1_i} &=& \int_0^{\tau_0} d\tau \, W_i \left( \frac{1 \! + \! \frac{H'}{aH^2} \!  +5s - f_{\rm evo}}{k} +  \mathfrak{cot}_K(\chi) \frac{2 -5s }{aH }  \right) S_{\Theta} \, \mathfrak{j}_{\ell}^{\ka\prime}  \nonumber \\
\Delta_{\ell}^{\mathrm{D}2_i} &=& \int_0^{\tau_0} d\tau \, W_i \left(f_\text{evo} -3\right) \frac{aH}{k^{2}}S_{\Theta} \, \mathfrak{j}_{\ell}^{\ka} \nonumber \\
\Delta_{\ell}^{\mathrm{G}1_i} &=& \int_0^{\tau_0} d\tau \, W_i \, S_\Psi \, \mathfrak{j}_{\ell}^{\ka} \nonumber \\
\Delta_{\ell}^{\mathrm{G}2_i} &=& -\int_0^{\tau_0} d\tau \, W_i  \left(3+\frac{H'}{aH^2}  -f_{\rm evo} + k\, \mathfrak{cot}_K(\chi) \frac{2 -5s }{aH } \right) S_\Phi \, \mathfrak{j}_{\ell}^{\ka} \nonumber \\
\Delta_{\ell}^{\mathrm{G}3_i} &=& \int_0^{\tau_0} d\tau \, W_i \, \left( \frac{1}{aH} \right) S_{\Phi'} \, \mathfrak{j}_{\ell}^{\ka} \nonumber \\
\Delta_{\ell}^{\mathrm{G}4_i} &=& \int_0^{\tau_0} d\tau \, W_i^{\mathrm{G}4} \,  S_{\Phi+\Psi} \, \mathfrak{j}_{\ell}^{\ka} \nonumber \\
\Delta_{\ell}^{\mathrm{G}5_i} &=& \int_0^{\tau_0} d\tau \, W_i^{\mathrm{G}5} \, S_{(\Phi+\Psi)} \, \mathfrak{j}_{\ell}^{\ka\prime} ~.
\label{eq:delta_terms}
\end{eqnarray}
As in Ref.~\cite{DiDio:2013bqa}, the different contributions correspond to density (`Den'), Kaiser redshift-space distortions (`Red'), lensing (`Len'), Doppler (`D1', `D2') and terms depending on the gravitational potentials (`G1'-`G5'), respectively.
For compactness we have omitted the arguments $q$ for the transfer functions, $(\tau,q)$ for the source functions, $\chi$ for the generalized Bessel functions, and $\tau$ for selection and background functions.
The index $i$ refers to the redshift bin around reference redshift $z_i$ and $W_i$ is a normalized window function over the bin.
Note that we use the functions $\mathfrak{j}_{\ell}^{\ka}$, $\mathfrak{j}_{\ell}^{\ka\prime}$, $\mathfrak{j}_{\ell}^{\ka\prime\prime}$, $\mathfrak{sin}_{K}$, $\mathfrak{cot}_{K}$ defined in eq.~(\ref{eq:gen_jl})-(\ref{eq:cotKgen}).
For the integrated terms `Len', G4 and G5, we have defined the modified window functions
\begin{eqnarray}
W_i^\mathrm{L}(\tau) &=& - \int_0^\tau \!\! d\tilde{\tau} W_i(\tilde{\tau}) \left( \frac{2-5s}{2} \right) \frac{k\,\mathfrak{sin}_K(\chi-\tilde\chi)}{\mathfrak{sin}_K(\chi)\mathfrak{sin}_K(\tilde\chi)} \nonumber \\
W_i^{\mathrm{G}4}(\tau) &=&  \int_0^\tau \!\! d\tilde{\tau} W_i(\tilde{\tau})  k\, \mathfrak{cot}_K(\tilde\chi) (2 -5s) \\
W_i^{\mathrm{G}5}(\tau) &=& \int_0^\tau \!\! d\tilde{\tau} W_i(\tilde{\tau})k \left(1+\frac{H'}{aH^2} +  k\, \mathfrak{cot}_K(\tilde\chi)  \frac{2 -5s}{ aH} + 5s-f_{\rm evo} \right)_{\tilde{\tau}}~. \nonumber
\end{eqnarray}
The corresponding expressions presented in \cite{DiDio:2013bqa} are recovered by taking the flat space limit and integrating by parts $\Delta_{\ell}^{\mathrm{G}5_i}$ (neglecting boundary terms since they vanish as $\tau\to0$ and are unobservable for $\tau=\tau_0$) and redefining consistently $\Delta_{\ell}^{\mathrm{G}1_i}$ and $\Delta_{\ell}^{\mathrm{G}2_i}$ so that the sum of these transfer functions coincides with the form considered in \classgal{}.

\section{Luminosity distance}
\label{sec:distance}
We derive the luminosity distance for the metric~(\ref{metric}) to first order in perturbation theory. In any metric, the (angular diameter) distance is determined by the Sachs focussing equation~\cite{sachs,ns}
\be
\frac{d^2 D_A}{d\lambda^2} = - \left( \mathcal{R} + \left| \sigma \right|^2 \right) D_A
\ee
with final conditions
\be \label{final_conditions}
D_A \left( \lambda_o \right) =0 \qquad \text{and} \qquad \left. \frac{d D_A\left( \lambda \right)}{d\lambda}\right|_{\lambda = \lambda_o} =\left. n^\mu u_\mu \right|_{\lambda = \lambda_o} \equiv - \omega_o
\ee
where
\be
\mathcal{R} = \frac{1}{2} R_{\mu \nu} n^\mu n^\nu
\ee
and
$\sigma$ is the complex shear of the light bundle defined as
\be
\sigma = \frac{1}{2} g \left( \epsilon , \nabla_{\epsilon} n \right) \qquad \text{with} \quad \epsilon \equiv e_1 + i e_2 \, .
\ee
Here $e_1$ and $e_2$ are orthonormal vectors, which are normal to both $u$ and $n$ at the observer position and parallel transported along the light-like geodesic parametrized by $n$. They form the so called `screen' of the light bundle. The distances $\tilde D_A$ and $D_A$ of conformally related metric $ d\tilde s^2$ and $ds^2=(1+\bar z)^2d\tilde s^2$, respectively, are simply connected through
\be \label{conformal_distances}
\tilde D_A = \left( 1 + \bar z \right)^{-1} D_A \qquad \mbox{and } \qquad \tilde D_L = \left( 1 + \bar z \right) D_L\, .
\ee
Here we have used the Etherington's reciprocity relation~\cite{Etherington}, $D_L=(1+ z)^2D_A$.
It is known~\cite{Seitz:1994xf} that the light shear $\sigma$ vanishes for conformally flat spacetimes. This is trivially the case for $K=0$ at the background level. More interesting, it is well known  that also the $K\neq 0$ cases are conformally flat (see, e.g.~problem 5 in chapter~V of \cite{choquet-bruhat1977} or, more recently, \cite{Iihoshi:2007uz}). This means, that at linear order, we need to solve perturbatively
\be \label{sachs_simple}
\frac{d^2 D_A}{d\lambda^2} = -  \mathcal{R}  D_A
\ee
where, adopting the conformal metric $ds^2=-(1+2\Psi)d\tau^2 +(1-2\Phi)\ga_{ij}dx^idx^j$ we have,
\bea
\mathcal{R}&=&  K + \frac{1}{2}\nabla^2\Psi + \ddot \Phi + 2 \bn \cdot \bnabla \dot \Phi - 2K \delta n^r + \frac{1}{2}\bnabla^2 \Phi + \frac{1}{2}\left( \bn \cdot \bnabla \right)^2 \left( \Phi - \Psi \right)
\nonumber \\
&=&
 K - 2 K \delta n^r +  \frac{d^2 \Phi}{d\lambda^2} + \frac{1}{2\CHI^2} \Delta_\Omega \left( \Psi + \Phi \right) + \frac{\CHI'}{\CHI} \left( \dot \Psi + \dot \Phi \right) -  \frac{\CHI'}{\CHI} \frac{d}{d\lambda} \left( \Psi + \Phi \right) \, .
\eea
For the second equality we have used repetitively the chain rule~(\ref{chain_rule}). This we will be convenient to simplify the final expression.

At the background level eq.~(\ref{sachs_simple}) reduces to
\be
\frac{d^2 \bar D_A}{d\lambda^2} = -   \mathcal{\bar R} \bar D_A = - K \bar D_A
\ee
which is solved by
\be
\bar D_A = \CHI \left( \lambda_o - \lambda \right)  \, .
\ee
Then, at first order, we have obtain the following equation for $\delta D_A$:
\be
\frac{d^2 \delta D_A}{d\lambda^2} +  \mathcal{\bar R} \delta D_A = -  \bar D_A\delta R \, .
\ee
which is solved by
\be
\delta D_A =- \omega_o^{(1)} \CHI \left(  \lambda_o - \lambda \right)  - \int_\lambda^{\lambda_o} \!\! \CHI\left( \lambda_o - \lambda' \right)
\CHI \left( \lambda' - \lambda \right) \delta R \left( \lambda' \right)  d \lambda' \, ,
\ee
where $\omega_o^{(1)}$ denotes the first order part of $\omega_o$ defined in eq.~(\ref{final_conditions}).
We express the distance in terms of the conformal time $\tau$ instead of the affine parameter $\lambda$ using
\be
\frac{d \tau}{d\lambda} = n^0 = 1+ \delta n^0 \left( \lambda \right)
\ee
and expanding
\be
\CHI \left(  \lambda_o - \lambda  \right) = \CHI \left( r \right)\left[ 1 - \frac{\CHI' \left( r \right)}{\CHI \left( r \right)}  \int_\tau^{\tau_o} \delta n^0 \left( \tau' \right) d\tau' \right]\,.
\ee
With this we obtain
\be
D_A =  \CHI\left( r   \right) \left[ 1 - \omega_o^{(1)}   - \frac{\CHI'\left( r \right)}{\CHI \left( r \right) }  \int_\tau^{\tau_o} \delta n^0 \left( \tau' \right) d\tau'
-\int_\tau^{\tau_o} \!\!
\frac{\CHI \left( r' \right)
\CHI \left( r' \right) }{\CHI\left(r \right)}\delta R \left( \tau' \right)  d \tau'
\right] \, .
\ee

The luminosity distance is related to the angular diameter distance through
\bea
D_L= \left( 1 + \delta z \right)^2 D_A &=& \CHI \left( r   \right) \left[ 1 +2 \delta z - \omega_o^{(1)}   -\frac{\CHI' \left( r \right)}{\CHI\left( r \right)} \int_\tau^{\tau_o} \delta n^0 \left( \tau' \right) d\tau'
  \right.
  \nonumber \\
  &&
  \left.
  - \int_\tau^{\tau_o} \!\!
  \frac{\CHI \left( r' \right)
    \CHI \left( r' \right) }{\CHI\left(r \right)}\delta R \left( \tau' \right)  d \tau'
  \right] \, .
\eea
Then, we include the expansion of the universe just considering the conformal metrics $ds^2$ and $ d\tilde s^2$, as shown in eq.~(\ref{conformal_distances}).
The last step consists in writing the luminosity distance in terms of the true observable redshift instead of  conformal time, or the unphysical background redshift. Following the approach of Ref.~\cite{Bonvin:2005ps} we compute
\be
\tilde D_L \left(\tau_s, \bn \right) = \tilde D_L \left( t \left( \bar z_s \right), \bn \right) \equiv \tilde D_L \left( \bar z_s , \bn \right)  = \tilde D_L \left( \tilde z_s , \bn \right) - \left. \frac{d}{d\tilde z} \tilde D_L \left( \tilde z , \bn \right) \right|_{\tilde z = \bar z } \delta \tilde z \, ,
\ee
with
\be
\left. \frac{d}{d\tilde z} \tilde D_L \left( \tilde z , \bn \right) \right|_{\tilde z = \bar z } = \frac{d}{d \bar z} \tilde D_L \left( \bar z , \bn \right) + \text{first order} = \frac{\tilde D_L}{1 + \bar z} \left( 1  + \frac{\CHI' \left( r \right)}{\CHI \left( r \right)}\frac{\left( 1 + \bar z \right) }{\HH} \right) + \text{first order} \, ,
\ee
where $\delta \tilde z = \left( 1 + \bar z \right) \delta z$.
This relates $\tilde D_L \left( \bar z , \bn \right)$ with $\tilde D_L \left( z , \bn \right) $, where from now  we denote the true observed redshift simply with z instead of $\tilde z $ and leads to
\bea
\tilde D_L \left( z, \bn \right) &=& \left( 1 + z \right)\CHI \left( r \right)  \left[
  1  + \delta z \left( 1 -\frac{\CHI' \left( r \right) }{ \HH \CHI \left( r \right)} \right) - \omega_o^{(1)}  -\frac{\CHI' \left( r \right)}{\CHI \left( r \right)} \int_\tau^{\tau_o} \delta k^0 \left( \tau' \right) d\tau'
  \right.
  \nonumber \\
  &&
  \left.
  - \int_\tau^{\tau_o} \!\!
  \frac{\CHI \left( r' \right)
    \CHI \left( r' \right) }{\CHI\left(r \right)}\delta R \left( \tau' \right)  d \tau'
  \right] \, .
\eea
To write  the luminosity distance explicitly in terms of the metric perturbations and peculiar velocities, we use
\bea
\delta z &=& - \left( \Psi + \bn \cdot \bv + \int_\tau^{\tau_o} \left( \dot \Psi + \dot \Phi \right) d\tau' \right) \, ,
\\
\omega_o^{(1)} &=& - \delta n^0_o \, ,
\\
\int_\tau^{\tau_o} \delta n^0 \left( \tau' \right) d\tau' &=&  \left( \tau_o -\tau \right) \delta n^0_o - \int_\tau^{\tau_o}d\tau' \int_{\tau'}^{\tau_o} \left( \dot \Psi + \dot \Phi \right) d\tau'' - 2 \int_\tau^{\tau_o} \Psi d\tau' \, ,
\\
\delta n^r &=& \delta n^r_o - \int_\tau^{\tau_o} \left( \partial_r \left( \Phi - \Psi \right) - 2 \dot \Phi \right) d\tau' = \delta n_o^r + \int_\tau^{\tau_o} \left( \dot \Psi + \dot \Phi \right) d\tau'  - \left( \Phi - \Psi \right) \;.
\nonumber \\
\eea
After some simplifications we then obtain the following expression for the perturbed luminosity distance in a spatially curved Friedmann Universe:
\bea
\tilde D_L \left( z, \bn \right)&=& \left( 1 + z \right) \CHI \left( r \right)
\left[
  1 - \left( \Psi + \bn \cdot \bv + \int_\tau^{\tau_o} \left( \dot \Psi + \dot \Phi \right) d\tau'\right)\left( 1 - \frac{\CHI' \left( r \right) }{ \HH \CHI \left( r \right)} \right)
  \right.
  \nonumber \\
  &&
  \left.
 + \frac{\CHI' \left( r \right)}{\CHI \left( r \right)} \int_\tau^{\tau_o} \left( \Psi + \Phi \right) d\tau' - \int_\tau^{\tau_o} \frac{\CHI \left( r -r' \right) }{\CHI \left( r \right) \CHI \left( r' \right) } \Delta_\Omega \left( \frac{\Psi + \Phi}{2} \right) d\tau' - \Phi
 \right] \, .
\nonumber \\
\eea

\section{Forecast specifications}
\label{sec:forecast}
In this section we summarize our forecast specifications and techniques.
We assume FLRW background described by the following fiducial values $\Omega_{\rm b}=0.05$, $\Omega_{\rm cdm}=0.25$, $\Omega_K=0$, $w_0=-1$ for the baryon, cold dark matter and curvature parameters, dark energy equation of state and sound speed, respectively. The dimensionless Hubble parameter is set to $h=0.67$.
All other standard $\Lambda$CDM cosmological parameters are set to values consistent with Planck \cite{Planck:2015xua} (however we neglect neutrino masses).
When varying the curvature parameter, we re-scale the dark energy one (when considering a cosmological constant, we also use the notation $\Omega_{\Lambda}$ instead of $\Omega_{\rm de}$) according to
\be
\label{eq:fried_constr}
\Omega_{\rm de}=1-\Omega_K-\Omega_{\rm m} \;,
\ee
where $\Omega_{\rm m}=\Omega_{\rm b}+\Omega_{\rm cdm}$,
which is consistent with neglecting the neutrino masses.

\begin{figure}[htb!]
  \includegraphics[width=0.45\linewidth]{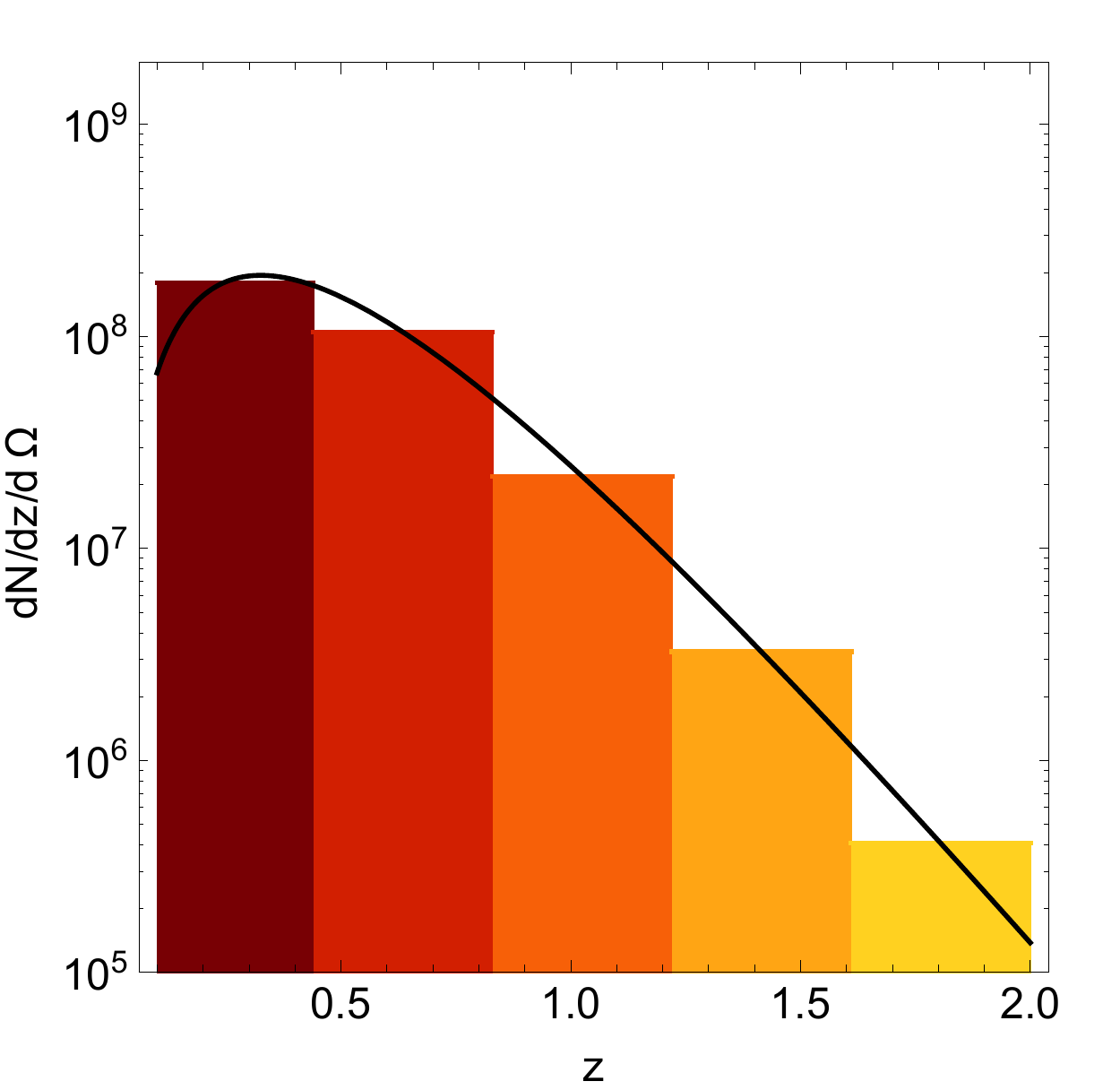}
  \quad
  \includegraphics[width=0.45\linewidth]{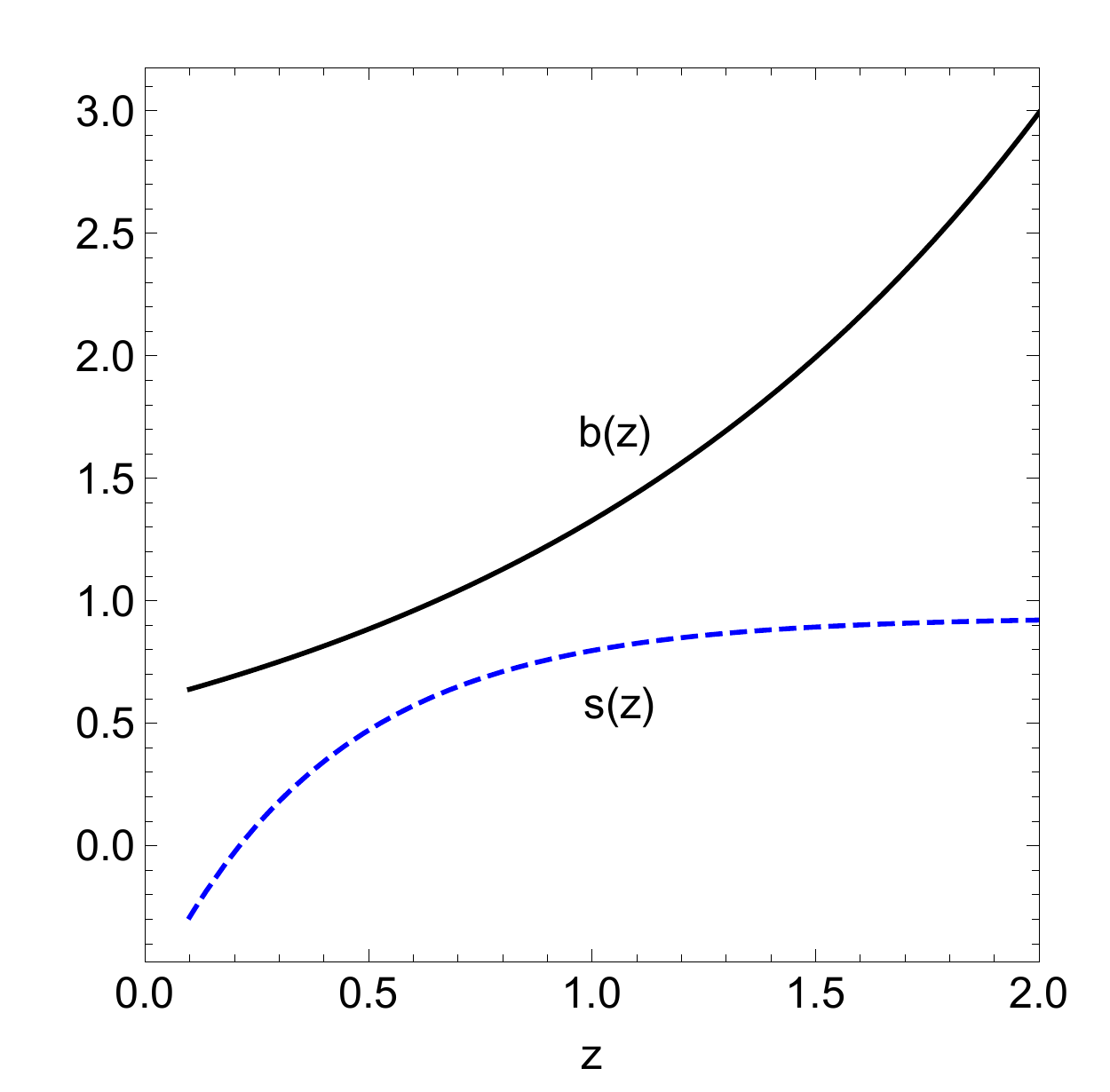}
  \caption{\emph{Left:} redshift distribution of galaxies for $0.05<z<2$; except when stressed differently, we group galaxies into 5 redshift bins of equal width, assuming a good redshift resolution.
    \emph{Right:} galaxy (solid) and magnification (dashed) biases.
  }
\label{fig:SKA_spec}
\end{figure}
We assume specifications consistent with a SKA-like survey as presented in~\cite{Santos:2015hra,Camera:2014bwa}.
In figure \ref{fig:SKA_spec} we summarize our specifications \cite{Santos:2015hra,Camera:2014bwa,Montanari:2015rga}:
\bea
&&\frac{dN}{dzd\Omega} = \left(\frac{180}{\pi}\right)^2 10^{c_1} z^{c_2}\exp\left( - c_3 z \right)\; \\
&&\quad \mbox{for} \quad 0.05<z<2.0\;, \nonumber \\
&&f_{\rm sky}=0.73\;,\\
&&b(z)= c_4 \exp\left( c_5 z \right) \;,\\
&&s(z)= c_6 + c_7 \exp\left( -c_8 z \right) \;, \label{sz_SKA}
\eea
where $c_1=6.7767$, $c_2=2.1757$, $c_3=6.6874$, $c_4=0.5887$, $c_5=0.8130$, $c_6=0.9329$, $c_7=-1.5621$, $c_8=2.4377$.
The number of galaxies per redshift and per steradian is indicated by $dN/dz/d\Omega$.
We divide the total redshift range in 5 redshift bins.
The sky coverage of the survey is given by $f_{\rm sky}$.
The magnification bias is consistent with the $5\;\mu$Jy sensitivity \cite{Camera:2014bwa}.
Given the spectroscopic redshift determination, we use tophat redshift bins.
We assume constant galaxy and magnification bias within each redshift bin.

To estimate error contours, we compute a series of angular auto- (within the $i^{th}-i^{th}$ bin) and cross- ($i^{th}-j^{th}$ , for $i\neq j$) power spectra, $C_{\ell} (z_i,z_j)$, in 5 different redshift bins.
We evaluate the covariance as (see e.g.~\cite{Asorey:2012rd,DiDio:2013sea}):
\begin{equation}
\label{eq:err-clgt}
{\rm Cov}_{C_{\ell \, \rm[(ij), (pq)]}} = \frac{\tilde{C}_{\ell}^{\rm (ip)} \tilde{C}_{\ell}^{\rm (jq)} + \tilde{C}_{\ell}^{\rm (iq)} \tilde{C}_{\ell}^{\rm (jp)}}{(2\ell+1)f_{\rm sky}},
\end{equation}
and the observed correlations include shot noise,
\begin{equation}
\tilde{C}_{\ell}  = C_{\ell}^{ij} + \frac{\delta_{ij}}{\mathcal{N}_i} \, ,
\end{equation}
where $\mathcal{N}_i$ denotes the number of sources per steradian in the $i^{th}$ bin.
An optimal binning in $\ell$-space, $\Delta\ell$, should be introduced to consider the covariance block-diagonal in multipoles \cite{Crocce:2010qi,Asorey:2012rd}, but we verified that it does not affect our results.
The Fisher matrix elements are given by:
\begin{equation}
F_{\alpha\beta} =
\sum_{\ell} \sum_{(ij)(pq)} \frac{\partial C_\ell^{ij}}{\partial \theta_\alpha}
\frac{\partial C_\ell^{pq}}{\partial
\theta_\beta} {{\rm Cov}_{C_{\ell \, \rm[(ij), (pq)]}}^{-1}} \, ,
\label{eq:Fisher}
\end{equation}
where $\theta_{\alpha(\beta)}$ is the $\alpha(\beta)$-th cosmological parameter, and we sum over $\ell$ up to 300 (unless stated differently).
The second sum is over the matrix indices $(ij)$ with $i \leq j$ and $(pq)$ with $p \leq q$ which run from 1 to the total number of bins (when all bin auto- and cross-correlations are taken into account).
Note that, when neglecting some of the bin correlations (e.g., neglecting cross-correlations), the covariance matrix must be first reduced to include only the correlations $(ij)$, $(pq)$ which are considered in the Fisher matrix (e.g., only $i=j$ and $p=q$ if cross-correlations are neglected) and then the resulting covariance must be inverted.

We assume that the fiducial Universe is described by the full relativistic spectra $C_{\ell}^{\rm rel}$.
Assuming the approximate Newtonian formalism and $C_{\ell}^{\rm nwt}$ spectra, where only density and redshift space distortion terms are included, may lead to a bias in the error estimation, as well as in the best-fit values.
We estimate the shift in the best-fit values due to the wrong model assumption as in \cite{Knox:1998fp,Heavens:2007ka,Kitching:2008eq,Camera:2014sba}.
We define the systematic error as $\De C_{\ell}=C_{\ell}^{\rm rel}-C_{\ell}^{\rm nwt}$.
Then the statistical bias on the best-fit is given by:
\begin{equation}
\label{eq:shift}
\Delta_{\theta_{\alpha}}=\sum_{\beta} \left[\left(F^{\text{nwt}}\right)^{-1}\right]_{\alpha\beta} B_{\beta} \;,
\end{equation}
where $F^{\text{nwt}}$ is obtained by replacing the full relativistic spectra in eq.~(\ref{eq:Fisher}) by the Newtonian ones $C_{\ell}^{\rm nwt}$, and:
\begin{equation}
B_{\beta} = \sum_{(ij)(pq)} \sum_{\ell} \De C_{\ell}^{ ij} \frac{\partial C_{\ell}^{{\rm nwt}\ pq}}{\partial \theta_{\beta}} {\rm Cov}_{C^{\rm nwt}_{\ell \, \rm[(ij), (pq)]}}^{-1} \;.
\end{equation}
This approximation neglects the fact that the systematical error also affects the covariance ${\rm Cov}_{C^{\rm nwt}_{\ell \, \rm[(ij), (pq)]}}$.
Furthermore, it has also been assumed that the bias is small compared to the marginal errors.
This in general not the case as correlations between redshift bins can be dominated by integrated terms like the lensing convergence.

\bibliographystyle{JHEP}
\bibliography{biblio_curvature}

\end{document}